\documentclass[journal=mamobx,manuscript=article]{achemso}

\usepackage[version=3]{mhchem} 
\usepackage[T1]{fontenc}       
\usepackage{siunitx}
\usepackage{epstopdf}
\usepackage{amssymb,amsmath}
\usepackage{xcolor}
\setkeys{acs}{usetitle = true}
\usepackage{caption}
\usepackage{pdflscape}

\graphicspath{{graphics/}}

\renewcommand{\vec}[1]{\boldsymbol{#1}}
\newcommand*{\citen}[1]{%
  \begingroup
    \romannumeral-`\x 
    \setcitestyle{numbers}%
    \cite{#1}%
  \endgroup   
}

\author{Maud Formanek}
\affiliation{Centro de F\'isica de Materiales (CSIC, UPV/EHU) and Materials Physics Center MPC, Paseo Manuel de Lardizabal 5, E-20018 San Sebasti\'an, Spain}

\author{Angel J. Moreno}
\affiliation{Centro de F\'isica de Materiales (CSIC, UPV/EHU) and Materials Physics Center MPC, Paseo Manuel de Lardizabal 5, E-20018 San Sebasti\'an, Spain}
\alsoaffiliation{Donostia International Physics Center (DIPC), Paseo Manuel de Lardizabal 4, E-20018 San Sebasti\'an, Spain}

\email{angeljose.moreno@ehu.es}

\title[SCNPs under shear]
  {Crowded Solutions of Single-Chain Nanoparticles under Shear Flow}

\abbreviations{SCNP, MPCD, MD, VMD}

\begin{document}

\newpage

\begin{abstract}
Single-chain nanoparticles (SCNPs) are ultrasoft objects obtained through purely intramolecular cross-linking of single polymer chains. By means of computer simulations with implemented hydrodynamic interactions, we investigate for the first time the effect of the shear flow on the structural and dynamic properties of SCNPs in semidilute solutions. We characterize the dependence of several conformational and dynamic observables on the shear rate and the concentration, obtaining a set of power-law scaling laws.
The concentration has a very different effect on the shear rate dependence of the former observables in SCNPs 
than in simple linear chains.
Whereas for the latter the scaling behavior is marginally dependent on the concentration, two clearly different scaling regimes are found for the SCNPs below and above the overlap concentration. 
At fixed shear rate SCNPs and linear chains also respond very differently to crowding. Whereas, at moderate and high Weissenberg numbers the linear chains swell, the SCNPs exhibit a complex non-monotonic behavior.
These findings are inherently related to the topological interactions preventing concatenation of the SCNP loops, leading to less interpenetration than for linear chains.

\end{abstract}

\newpage

\section{I. Introduction} 
 
Single-chain nanoparticles (SCNPs) are synthesized through purely intramolecular 
bonding of functionalized polymer chains \cite{PomposoSCNPbook}. These fully polymeric nano-objects are the basis of the single-chain technology, a rapidly growing research area due to the recent advances demonstrating their promising application in fields so diverse as catalysis, drug delivery, biosensing or nanocomposite design
\cite{terashima2011,perez2013endowing,huerta2013,tooley2015,hamilton2009,sanchez2013design,gillisen2012single,Mackay2003,Bacova2017}. Taking inspiration from biological systems such as proteins or enzymes, it is a long term goal to design SCNPs with precise control over the chemical sequence and molecular architecture, high performance and quick response to environmental changes.
Much of the research on SCNPs has been devoted to advanced synthesis and to the implementation of enhanced functionalities (catalytic, luminiscence, etc). Comparatively, little is still known about their physical properties (structure and dynamics). This is a key question, since the functionality of SCNPs should be in part related to their internal structure and dynamics (allowing, e.g., for fast response to changes in pH or temperature
and for adaptation to multiple substrates). Moreover, structure and dynamics can be strongly altered in situations such as flow, confinement or crowding that are ubiquitous in multiple problems of practical interest, as e.g., diffusion in blood, membranes or cell environments.
A series of investigations by small-angle X-ray and neutron scattering \cite{Moreno2013,Basasoro2016,Arbe2016,GonzalezBurgos2018} have revealed that the molecular topology of SCNPs obtained through conventional routes is far from a compact, globular nano-object \cite{Pomposo2014a,Pomposo2017rever}. In the good solvent conditions where the synthesis is performed, 
the linear precursors universally adopt self-avoiding random-walk conformations \cite{Rubinstein2003}. These conformations strongly promote bonding of reactive groups that are separated by short contour distances. On the contrary, two groups separated by long contour distances are statistically far from each other in the real space, and are unlikely to form cross-links.  Though such events occur, their number is very low (the probability decays as a power-law with the contour distance) and insufficient to fold the precursor into a compact object \cite{Moreno2013,LoVerso2014,Rabbel2017,Formanek2017,Oyarzun2018,Moreno2018}. 
For a fixed length and fraction of reactive groups in the precursor the obtained SCNPs are topologically polydisperse \cite{Moreno2013,LoVerso2014,Formanek2017,Moreno2018}. Thus, each initial stochastic realization of the precursor leads to a different network structure of the resulting SCNP. A distribution of network topologies is formed, 
which is largely dominated by sparse structures \cite{Moreno2013,Moreno2016JPCL}. SANS experiments and simulations have revealed some structural similarities between intrinsically disordered proteins (IDPs) and SCNPs \cite{Moreno2016JPCL}. Though the latter lack the ordered regions present in IDPs, they still contain weakly deformable compact domains connected by flexible strands, suggesting that SCNPs in concentrated solutions can be used as model systems, free of specific interactions, to shed light on the effect of excluded volume on IDPs in crowded environments. 

The particular internal structure of SCNPs, containing loops and clusters of loops of different sizes, has a very different response in solution
when the concentration is increased above the overlap density and up to the melt state. Whereas linear chains show a crossover from self-avoiding to Gaussian conformations, SCNPs collapse to more compact conformations \cite{Moreno2016JPCL,GonzalezBurgos2018,Oberdisse2019} resembling those of the so-called fractal or `crumpled' globule \cite{Grosberg1988,Mirny2011}, characterized by a loose core and outer protrusions. As a consequence of the topological interactions (non-concatenation of the loops), the SCNPs in concentrated solutions and melts show a weaker interpenetration than linear chains and some microsegregation in close analogy to ring polymers, which have been invoked as model systems to explain the formation of chromosome territories \cite{Halverson2014}.

Very recently, we have investigated non-equilibrium aspects of isolated SCNPs (mimicking high dilution), namely the effect of shear flow on their structural and dynamic properties \cite{Formanek2019}. We have characterized the dependence of several observables of interest measuring the size, orientation and intrinsic viscosity, on the applied shear rate. The obtained power-laws have characteristic exponents that are clearly different from those found in other architectures (linear chains, rings, stars, dendrimers \cite{Aust1999,Schroeder2005,Ripoll2006,Nikoubashman2010,Chen2013,Chen2013SM,Chen2015,Chen2017,Liebetreu2018,Jaramillo-Cano2018,Formanek2019}). Thus, SCNPs constitute a novel class of macromolecules with distinct response to shear. Interestingly, this response is, at most, weakly dependent on the specific topology of the SCNP, and it seems inherently related just to its network-like architecture \cite{Formanek2019}. 

Most of the studies on the conformations and intramolecular dynamics of polymeric systems under shear flow 
have been performed at high dilution (experimental) or for isolated polymers (simulations).
Only a few studies in linear chains and star polymers have addressed the effect of the concentration
\cite{Hur2001,Huang2010mac,Huang2011,Huang2012jpcm,Fedosov2012,Singh2013}. 
In this article we investigate, for the first time, the structural and dynamic properties of semidilute solutions of SCNPs under shear flow. We employ large-cale simulations including hydrodynamic interactions.
We characterize the dependence of several conformational and dynamic observables on the shear rate and the concentration. We find that, when compared to simple linear chains, SCNPs exhibit a very different response to shear and crowding. Whereas linear chains esentially show a single power-law dependence on the shear rate, SCNPs exhibit two distinct regimes, with a crossover around the
overlap concentration. At fixed shear rate, the size of the SCNPs shows a complex dependence on the concentration. Whereas crowding at fixed moderate and high shear rate leads to swelling of linear chains, the SCNPs may show both swelling and shrinking, as well as reentrant behavior. These findings are inherently related to the topological interactions preventing concatenation of the SCNPs, which lead to less interpenetration than for linear chains. 

\section{II. Model and Simulation Details} 
 
The simulated SCNPs were based on the bead-spring model with purely repulsive interactions \cite{Kremer1990}, capturing the basic ingredients of the system: monomer excluded volume, connectivity, and chain uncrossability (which moreover prevents concatenation of the permanent loops of the SCNPs). The SCNPs were generated trough irreversible
cross-linking of isolated (mimicking the limit of high dilution) linear precursors of $N=200$ monomers, of which a 25\%  were reactive groups randomly distributed along the chain contour, with the condition of not being consecutively placed to avoid trivial cross-links. A total of 200 fully reacted SCNPs were used for the simulations of the solutions. The generated SCNPs were topologically polydisperse (see typical equilibrium conformations of different SCNPs at high dilution 
in Figure~S1 of the Supporting Information (SI)). Though some of them were relatively compact `nanogel-like' networks, most of them were sparse objects \cite{Formanek2019}. Two kind of solutions were investigated: i) topologically polydisperse, where different SCNPs were taken from the generated set and were placed in the simulation box, ii) topologically monodisperse, where all the SCNPs were replicas of the same one. Three monodisperse solutions were investigated, formed by SCNPs with a low, middle and high asphericity parameter, at the extremes and center of the obtained distribution of equilibrium asphericities \cite{Formanek2019}. 
We use the indices $x,y,z$ to denote the directions of the flow, gradient and vorticity, respectively (see setup in Figure~S2 in the SI).
A linear shear profile was imposed by Lees-Edwards boundary conditions \cite{Lees1972}. The hydrodynamic interactions were implemented through the multi-particle collision dynamics (MPCD) technique \cite{Malevanets1999}.
Further details about the model and the simulation methods are given in the SI
and in Ref.~\citen{Formanek2019}.

If $R_{\rm g} = \langle R^2_{\rm g} \rangle^{1/2}$ is the radius of gyration at equilibrium (zero shear rate), 
we define the overlap density as $\rho^{\star}= N(2R_{\rm g})^{-3}$, i.e., 
as the number density of a cube of size $2R_{\rm g}$ containing the $N$ monomers of a SCNP. For concentrations higher than $\rho^{\star}$ the clouds of monomers of the surrounding macromolecules enter in the cube, distorting the conformations with respect to dilute conditions. Linear chains and SCNPs experience a crossover to Gaussian and crumpled globular conformations, respectively {\cite{Moreno2016JPCL,GonzalezBurgos2018}.   
In what follows the concentration of the solution, $\rho = N_{\rm m}/V$, with $N_{\rm m}$ the total number of monomers in the simulation box and $V$ the volume of the box, will be given in reduced units, $\rho/\rho^{\star}$. We explored concentrations in the range $0.25 \leq \rho/\rho^{\star} \leq  6.24$. The highest concentration corresponds to a monomer density $\rho = 0.38$, qualitatively corresponding to 300-400 mg/mL \cite{Moreno2016JPCL}. The SCNPs are unentangled even at the highest concentration. For linear chains of the same $N=200$ in good solvent the entanglement concentration is $\rho_{\rm e} \approx (N_{\rm e}/N)^{3\nu_{\rm F} -1}$ with $N_{\rm e}$
the entanglement length in the melt and $\nu_{\rm F} = 0.59$ the Flory exponent \cite{Rubinstein2003}. Since for the used bead-spring model
$N_{\rm e} \gtrsim 65 $ \cite{Ever_science,Kroger_Ne}, the entanglement concentration is  $\rho_{\rm e} \gtrsim 0.42$, above the highest simulated concentration of SCNPs. For the SCNPs, which are less penetrable than linear chains,
a reduction of entanglements with respect to their linear counterparts is expected
\cite{Arbe2019meso}, so that their $\rho_{\rm e}$ will be even higher.

We explored shear rates in the range $5\times 10^{-5} \leq \dot{\gamma} \leq 2 \times 10^{-2}$. In the rest of the article the shear rates will be given in units of the dimensionless Weissenberg number, $Wi = \dot{\gamma}\tau$, where $\tau$ is the relaxation time at equilibrium and high dilution $\rho =0$. The value of $\tau$ was determined from the exponential decay of the correlator of $R_{\rm g}$ \cite{Formanek2019}. 
For low Weissenberg numbers $Wi \ll 1$ the characteristic time for intramolecular relaxation is much shorter than the characteristic time of the shear flow, and the conformations are weakly perturbed with respect to equilibrium. For $Wi \gg 1$ the macromolecule cannot relax their conformations in the fast flow and is strongly elongated most of the time, though it may experience more compact transient conformations due to tumbling motion \cite{Dalal2012,Lang2014,Formanek2019},
where the polymer contracts, flips around and extends again, with the head and tail having switched sides.

\section{Results}

\begin{figure}[ht]
\centering
\includegraphics[width=0.85\linewidth]{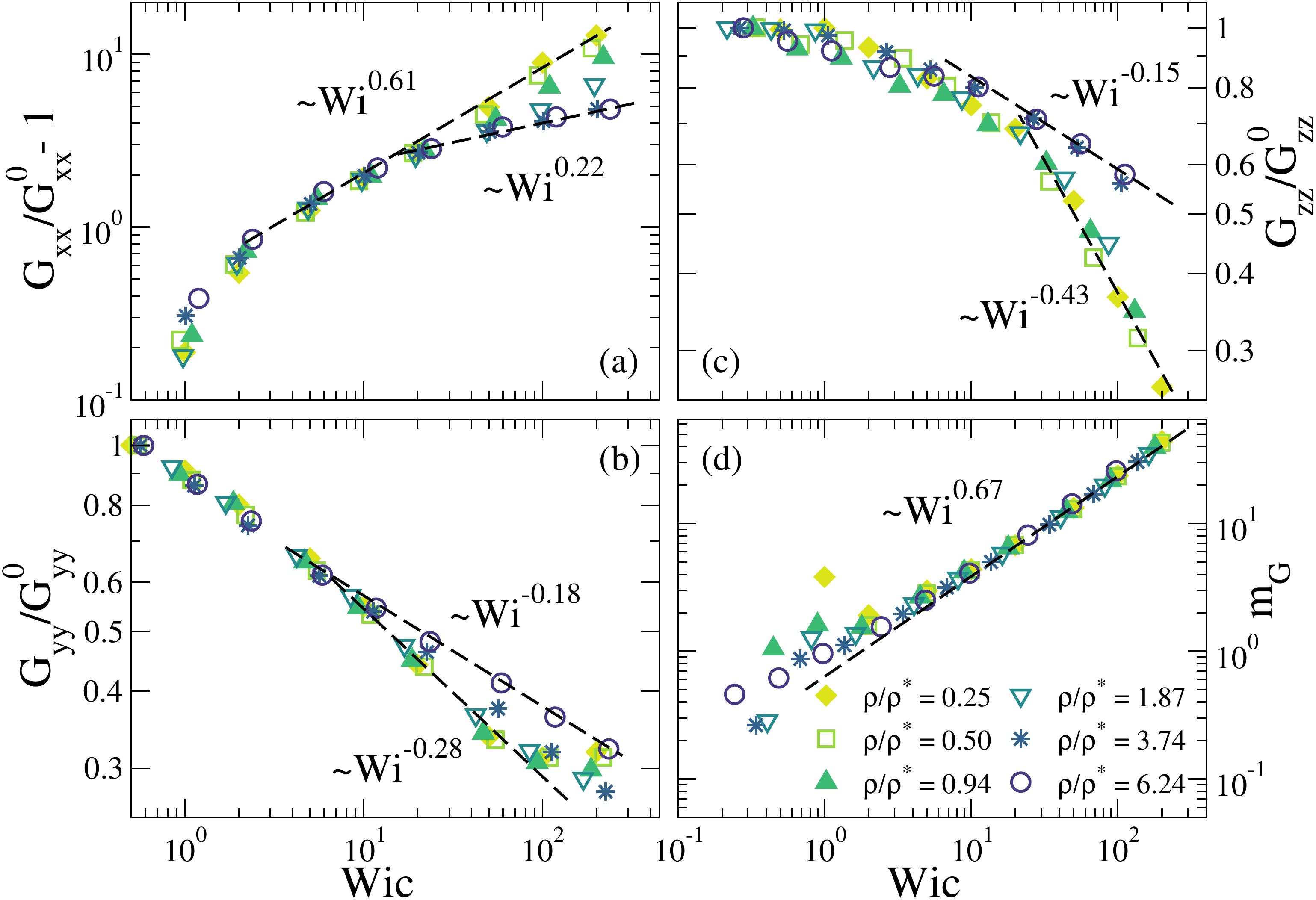}
\caption{For the SCNPs in the polydisperse solutions,
diagonal components of the inertia tensor (a-c) and orientational resistance (d) vs. the rescaled Weissenberg number. 
Each data set corresponds to a fixed concentration (see legend). The components of the inertia tensor are normalized
by their values at their corresponding concentration and the lowest simulated shear rate.
Dashed lines represent power laws.}
\label{fig:poly-Gallmg-weiss}
\end{figure}

We start our analysis by characterizing static observables adapted to the geometry of the shear flow. 
The panels (a-c) of Figure~\ref{fig:poly-Gallmg-weiss} show the $Wi$-dependence of the diagonal components $G_{\mu\mu}$ of the gyration tensor, along the flow ($x$), gradient ($y$) and vorticity ($z$) directions, in the topologically polydisperse solutions. The gyration tensor is computed as
\begin{equation}
G_{\mu\nu} =\frac{1}{N} \sum_{i=1}^{N}  (r_{i,\mu}-r_{{\rm cm},\mu})(r_{i,\nu}-r_{{\rm cm},\nu}) \, , 
\end{equation}
where $r_{i,\mu}$ and $r_{{\rm cm},\mu}$ are the $\mu$-th Cartesian components of the position of monomer $i$ and the center-of-mass of the SCNP respectively.
Each data set corresponds to a fixed value of the normalized concentration $\rho/\rho^{\star}$, and the data have been normalized by the values, $G_{\mu\mu}^0$, at such a concentration and the lowest simulated shear rate $\dot{\gamma}=5\times 10^{-5}$. The panel (d) shows the corresponding data sets for the $Wi$-dependence of the orientational resistance $m_{\rm G}$ \cite{Bossart1995}. This is defined as $m_{\rm G} = Wi \tan(2\theta) = 2Wi G_{xy}/(G_{xx}-G_{yy})$,
where $\theta$ is the angle between the direction of the largest eigenvector of the gyration tensor and the 
direction of the flow. Thus, for a fixed $Wi$ lower values of $m_{\rm G}$ mean stronger alignment with the flow.
In all panels each $Wi$ has been rescaled by a factor to obtain the best overlap with the data set at the lowest simulated concentration $\rho/\rho^{\star} = 0.25$ (the factor is 1 for this concentration). This representation as a function of the rescaled Weissenberg number ($Wic$) is made to highlight the emergence of master curves and scaling behavior. A remarkable feature is observed in the components of the gyration tensor: whereas at low and moderate shear rates a single scaling is apparently observed, at high rates ($Wi \gg 1$)  two clearly different power-law scaling regimes are found for low ($\rho/\rho^{\star} \ll 1$) and high ($\rho/\rho^{\star} \gg 1$) concentration. This observation is rather different from the case of linear chains \cite{Huang2010mac,Huang2012jpcm}.
In these systems increasing the density even far beyond the overlap concentration has, at most, a very weak effect in the $Wi$-dependence of the $G_{\mu\mu}$ components, which esentially keep the power laws found at dilute conditions. In the SCNPs the crossover between the low and high concentration scaling regimes takes place in a different concentration regime for each component of the gyration tensor. These differences might be due to the asymmetric change in shape under shear flow, which leads 
the polymers to effectively overlap at different concentrations in different directions. Still,
in all cases the crossover is found at concentrations of the order of the equilibrium overlap density. 
The results in panels (a-c) reveal the strong effect of crowding on the scaling of the SCNP size under shear. However, crowding has little or no effect on the $Wi$-dependence of the orientation in shear flow. As can be seen in  Figure~\ref{fig:poly-Gallmg-weiss}d, data for $m_{\rm G}$ at all the concentrations are consistent with the same power law,
i.e., the molecular orientation of the inertia ellipsoid reacts to shear in the same way, irrespective of the specific effect of crowding on the molecular size and shape.

\begin{figure}[ht]
\centering
\includegraphics[width=0.46\linewidth]{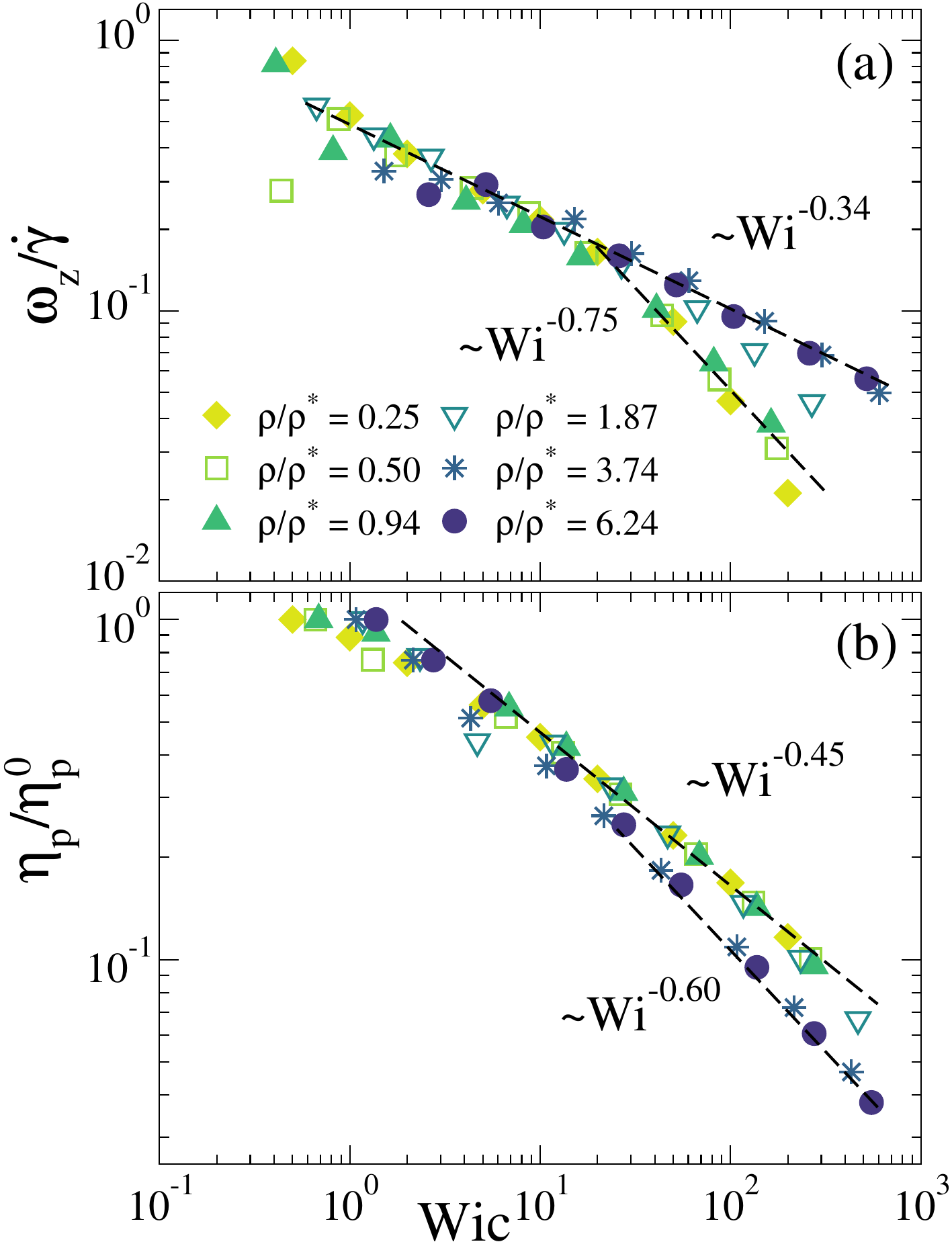}
\caption{As Figure~\ref{fig:poly-Gallmg-weiss} for the rotational frecuency scaled by $\dot{\gamma}^{-1}$ (a) and the
polymer contribution to the viscosity (b).}
\label{fig:poly-rotvisc-weiss}
\end{figure}

Panels (a) and (b) of Figure~\ref{fig:poly-rotvisc-weiss} show the $Wi$-dependence of the rotational frequency $\omega_z$ and the viscosity $\eta_{\rm p}$ 
(polymer contribution), 
respectively, for the polydisperse solution.
The rotational frequency has been determined by using the relation $\vec{\bf L} = {\bf J}\vec{\bf \omega_z}$,
where  $\vec{\bf L}$ and ${\bf J}$ are the angular momentum and inertia tensor, respectively. The polymer
contribution to the viscosity is obtained as $\eta_{\rm p} = \sigma_{xy}\dot{\gamma}^{-1}$, where $\sigma_{xy}$
is the $xy$-component of the Kramers-Kirkwood stress tensor \cite{Bird1987}:
\begin{equation}
\sigma_{\mu\nu} = -\sum_{i=1}^N \langle r_{i, \mu} F_{i, \nu}\rangle  . 
\end{equation}
$\vec{\bf F}_i$ is the total force acting on monomer $i$ and $\mu,\nu$ denote the Cartesian components. 
As in Figure~\ref{fig:poly-Gallmg-weiss}, each data set in Figure~\ref{fig:poly-rotvisc-weiss} corresponds to a fixed concentration, and the Weissenberg numbers are rescaled to obtain the best overlap with the data for $\rho/\rho^{\star} = 0.25$. We find the same qualitative behavior as for the components of the gyration tensor: data at low $Wi$ show the same scaling, whereas at high $Wi$ two different scaling regimes are found, and the crossover between both regimes takes place when $\rho$ is increased above the overlap concentration. The general trend for the diagonal components of the inertia tensor and the rotational frequency is to follow a weaker dependence on the shear rate at high concentrations (lower exponents). Thus, in crowded solutions shearing is less efficient for deforming and rotating the SCNPs than at high dilution, suggesting that deformation and rotation are hindered by the steric interactions with the surrounding crowders. The polymer contribution to the viscosity shows the opposite effect: shearing at high densities leads to a stronger reduction of $\eta_{\rm p}$. The number of side contacts at high concentration is large, so that stretching the SCNPs removes many more contacts and is more efficient  to reduce the viscosity than at lower concentrations.

Figures~S3 and S4 in the SI show, for the monodisperse solutions of SCNPs with low and high asphericity, respectively, typical snapshots of the simulation box for different concentrations and Weissenberg numbers. All the SCNPs in the solution are represented. The color codes are assigned according to the instantaneous value of $R_{\rm g}$. The snapshots for the polydisperse systems (not shown) display similar features. As can be seen, at high concentrations 
and moderate $Wi$ the SCNPs mantain the structural characteristics found in equilibrium. Due to the topological interactions 
that prevent concatenation of the loops, they adopt more compact conformations and are less interpenetrated than linear chains
\cite{Moreno2016JPCL}. At high values of the concentration and Weissenberg number there is some microsegregation between SCNPs with stretched and compressed instantaneous configurations. The qualitative picture of Figures~S3 and S4 rationalizes the existence of two scaling regimes (at low and high concentration)
for the $Wi$-dependence of the size and viscosity of the SCNPs, in contrast with the essentially single scaling independent of the concentration
found for linear chains. At high concentrations and in equilibrium the linear chains are strongly interpenetrated and their conformations are much less perturbed
with respect to high dilution. When the chains are sheared they are still weakly perturbed with respect to high dilution at the same $Wi$, since unlike for SCNPs, stretching is not limitted by weaker penetrability and non-concatenability. As a consequence, for the linear chains crowding has no significant effect in the $Wi$-dependence of the relative change of their molecular size and viscosity.

\begin{figure}[ht]
\centering
\includegraphics[width=0.46\linewidth]{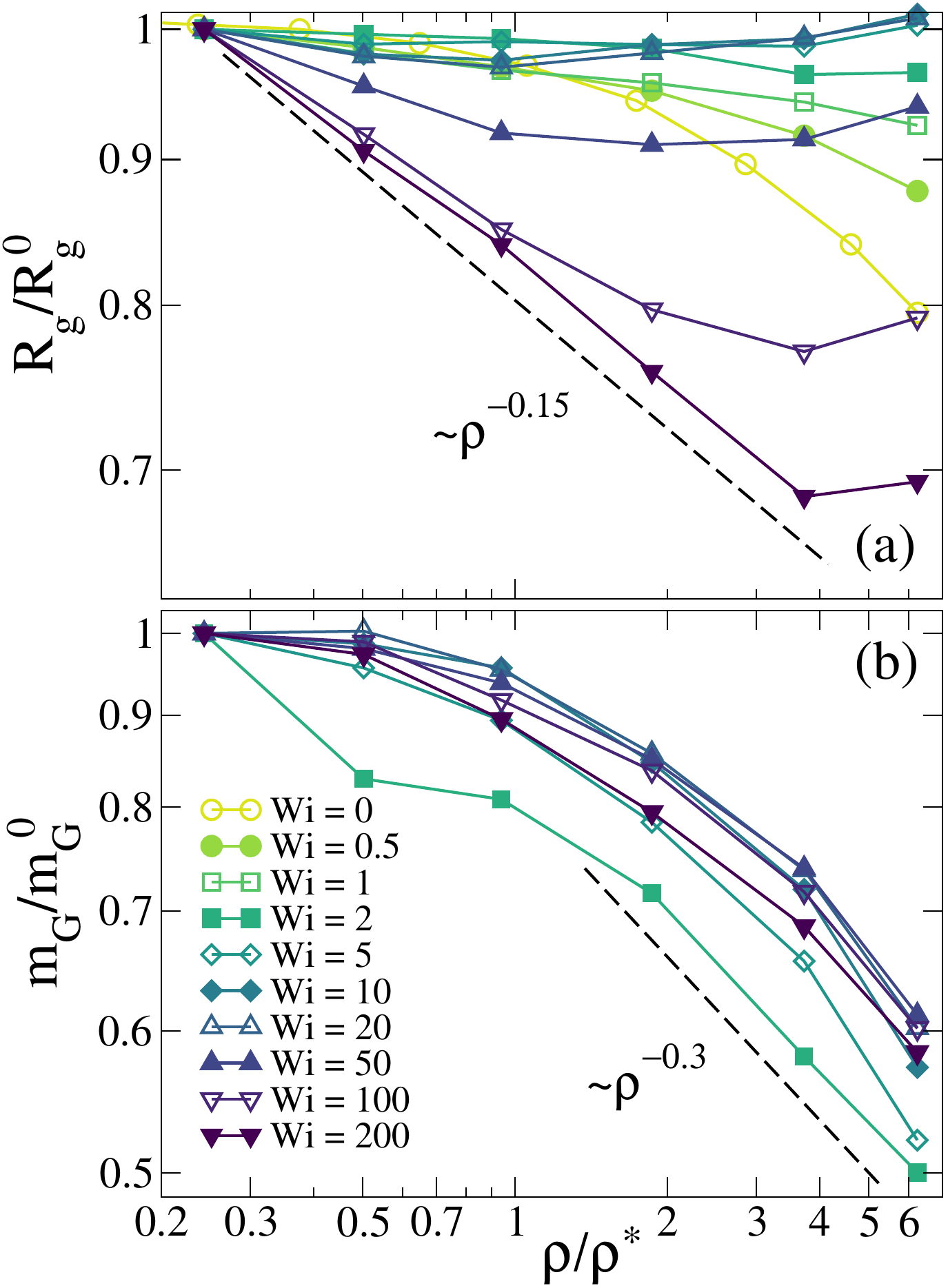}
\caption{For the SCNPs in the polydisperse solutions,
gyration radius (a) and orientational resistance (b) vs. the concentration. 
Each data set corresponds to a fixed Weissenberg number (see legend) and is normalized by the
value ($R_{\rm g}^0$, $m_{\rm G}^0$) at its corresponding $Wi$ and concentration $\rho/\rho^{\ast} = 0.25$.
Dashed lines represent power laws.}
\label{fig:poly-Rgmg-dens}
\end{figure}

Figure~\ref{fig:poly-Rgmg-dens} shows the gyration radius $R_{\rm g}$ and the orientational resistance $m_{\rm G}$
vs. the normalized concentration for the polydisperse solution. Figure~S5 in the SI shows analogous results for the components of the gyration tensor. In all cases each data set corresponds to a fixed value of the Weissenberg number, and it is normalized by its corresponding value  ($R^0_{\rm g}$, $m^0_{\rm G}$, $G^0_{\mu\mu}$) 
at  $\rho/\rho^{\star} = 0.25$. For any fixed Weissenberg number, increasing the concentration leads to a reduction
of the orientational resistance $m_{\rm G}$, i.e., the SCNPs tend to be more aligned with the flow as the solution becomes more crowded.
The data for $R_{\rm g}$ in Figure~\ref{fig:poly-Rgmg-dens}a reveals a much more complex behavior. As expected, increasing the concentration of the solution above the overlap density leads, in equilibrium ($Wi=0$), to shrinking of the SCNPs. This behavior is still found in the weakly and moderately sheared solutions ($Wi \leq 1$), though a much weaker shrinking is observed as $Wi$ is increased. For $1 \leq Wi \leq 20$ there is esentially no effect of the concentration: adding more SCNPs to the sheared solution, even up to $\rho/\rho^{\star} \sim 6$, does not change their mean size, or even leads to some weak swelling. Unlike at lower shear rates,
the SCNPs are, in average, sufficiently elongated to fill the space without significant contact with their neighbors even at high concentrations,
and their size is unaltered with respect to high dilution. 
This effect is partially reversed by further increasing the shear rate, for which a non-monotonic dependence of the molecular size on the concentration is found. At $Wi > 20$ adding more SCNPs to the solution leads to shrinking (with a stronger effect for higher $Wi$), but the SCNPs start to swell if the concentration is further increased.

Since the radius of gyration is given by $R_{\rm g}^2 = G_{xx}+G_{yy}+G_{zz}$, one expects that the scenario
displayed in Figure~\ref{fig:poly-Rgmg-dens}a for the SCNPs elongated under shear flow esentially comes from the largely
dominant $x$-contribution of the gyration tensor. This is confirmed by panel (a) of Figure~S5, where $G_{xx}$ shows all the qualitative trends found for $R_{\rm g}$. On the contrary, the component along the gradient direction, $G_{yy}$, monotonically shrinks with increasing concentration for all the Weissenberg numbers, which is consistent with the stronger alignment reflected in the behavior of the orientational resistance (Figure~\ref{fig:poly-Rgmg-dens}b). Crowding at low and moderate $Wi$
shrinks the molecular size along the vorticity direction $z$, as can be seen for $G_{zz}$ in panel (c) or Figure~S5. 
At high $Wi$ the behavior is non-monotonic, the SCNPs initially swell along the $z$-direction and above some concentration they start to shrink. As can be seen in panels (a) and (c) of Figure S5, $G_{xx}$ and $G_{zz}$ at fixed $Wi$ 
qualitatively show opposite dependences on the concentration. Thus, increasing the concentration leads to a stronger alignment with the flow and a redistribution of the monomers within the SCNP, through stretching along one of the $x,z$-directions and shrinking along the other one.

\begin{figure}[ht]
\centering
\includegraphics[width=0.40\linewidth]{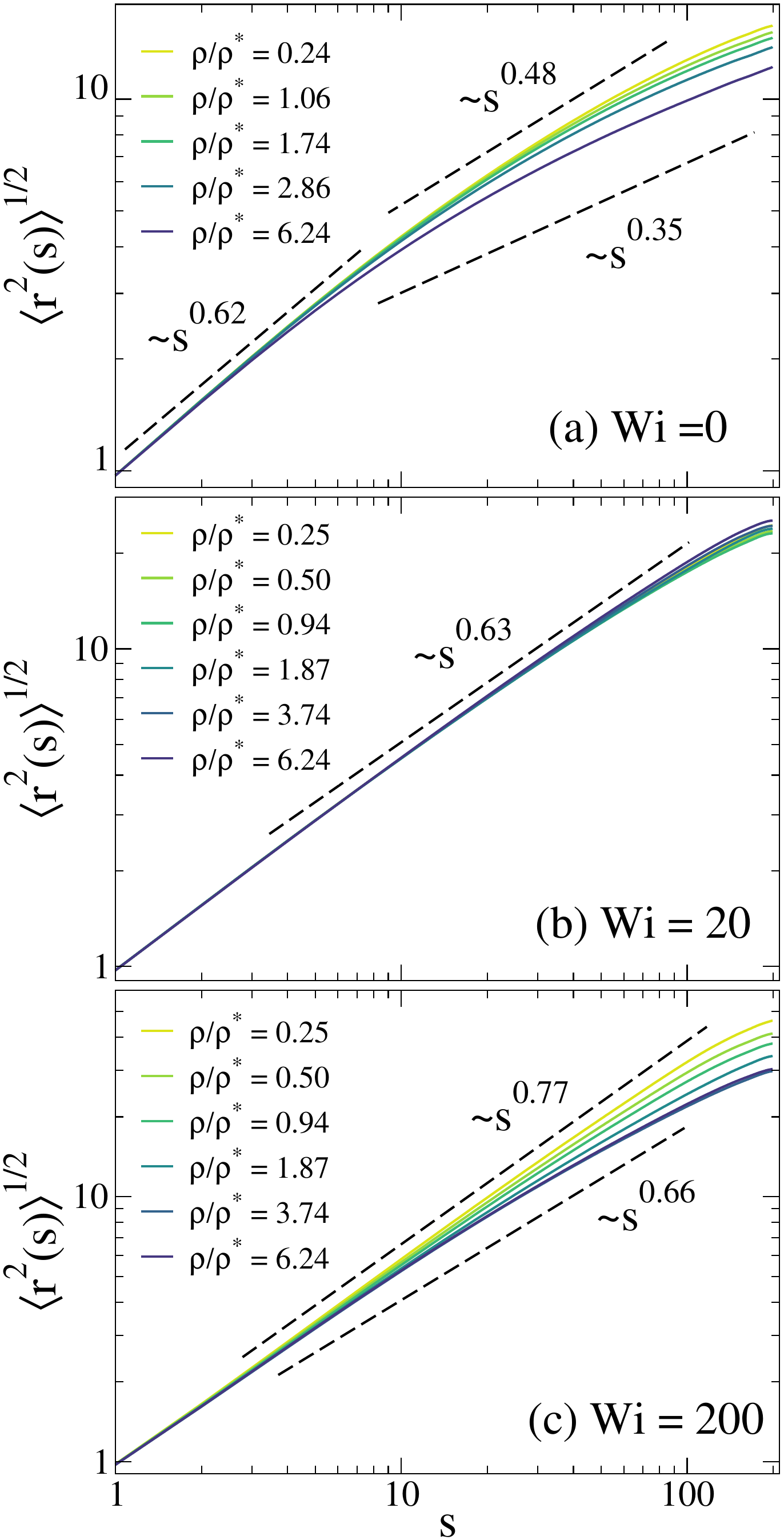}
\caption{For the SCNPs in polydisperse solutions, real vs intramolecular contour distance at fixed Weissenberg numbers
$Wi=0$ (a), $Wi=20$ (b) and $Wi=200$ (c). Each data set corresponds to a concentration (see legends). Dashed lines represent approximate power laws.}
\label{fig:poly-rs}
\end{figure}

It is worth mentioning that the emerging scenario displayed in Figure~\ref{fig:poly-Rgmg-dens}a is not related to a complex interplay of contributions of the different molecular topologies present in the polydisperse solution, responding in a different way to crowding under shear. Figure~S6 in the SI shows the corresponding results for the topologically monodisperse solutions. For the three (low, middle and high) asphericities investigated the same qualitative scenario is found and the differences are only quantitative. Not surprisingly, the most deformable SCNPs, i.e., those with the highest asphericity and most sparse structures, are more affected by crowding the solution (note the highest exponent in the approximate scaling $R_{\rm g} \sim \rho^{-\alpha}$ at $Wi=200$ in Figure~S6).

Further insight on the microscopic origin of the complex dependence of the SCNP size on concentration and shear rate can be 
obtained by analyzing their intramolecular correlations.
Figure~\ref{fig:poly-rs}  shows the real space distance $r(s)=\langle r^2 (s)\rangle^{1/2}$ 
vs. the contour distance $s$ in equilibrium ($Wi =0$) \cite{GonzalezBurgos2018} and for $Wi=20$ and 200. By labelling the monomers
as $i=1,2,...,N$ according to their position in the linear backbone of the precursor, the contour distance
is defined as $s = |i-j|$, and the real distance is just $r = |\vec{\bf r}_i - \vec{\bf r}_j|$.
The quantity $r(s)$ provides insight on the conformational statistics of the SCNPs, through the exponent $\nu$
of the scaling law $r(s) \sim s^{\nu}$. It should be noted that the investigated SCNPs of $N=200$
are not large enough to develop a well-defined power-law regime over a broad $s$-range. Moreover a significant 
fraction of SCNPs have some long loop of countour length $N/2 < l < N$ \cite{Moreno2016JPCL}. 
Obviously, by moving forward along the contour of such a loop the real distance $r(s)$ will stop growing at some point 
when the path starts to go back to the origin.
The contribution of the SCNPs containing such long loops rationalizes the observed flattening of $r(s)$ at large $s$.
At short scales ($s < 10$) the SCNPs in equilibrium ($Wi=0$, panel (a)) show a scaling exponent $\nu \sim 0.6$ similar to the Flory exponent for self-avoiding walks, indicating that at such scales the SCNPs effectively behave as linear chains with excluded volume interactions. 
The effect of the cross-links on the scaling of $r(s)$ becomes evident at larger distances. In dilute conditions
($\rho/\rho^{\star} = 0.25$) an exponent $\nu \sim 0.5$ is observed. This is similar to the exponent expected
for linear chains in $\theta$-solvent conditions ($\nu = 1/2$), where only local compaction occurs and the large-scale statistics
is that of a random-walk \cite{Rubinstein2003}. In the case of SCNPs in the simulated {\it good} solvent conditions this local compaction is mediated by a majority of cross-links between reactive groups close in the chain contour \cite{Moreno2013,Moreno2016JPCL}. By increasing the concentration above the overlap density a crossover to a lower exponent $\nu \sim 0.35$ is observed. This is rather different from the well-know transition 
in linear chains from the Flory  ($\nu_{\rm F} = 0.59$) to the Gaussian value ($\nu = 1/2$) \cite{Rubinstein2003}. The exponent found for the SCNPs
is similar to the value $\nu = 1/3$ for fractal globules \cite{Grosberg1988,Mirny2011}. 
For relatively large Weissenberg numbers, $Wi=20$, the chain statistics of the SCNPs is almost unaffected by the concentration
(Figure~\ref{fig:poly-rs}b). 
This is consistent with the very weak effect observed in the molecular size 
(see data for $Wi=20$ in Figure~\ref{fig:poly-Rgmg-dens}a). The exponent $\nu = 0.63$ indicates that the typical conformations
are more elongated than self-avoiding random walks ($\nu_{\rm F} =0.59$) but still very far from straight rods ($\nu_{\rm R} =1$). At the highest investigated Weissenberg number ($Wi=200$), rod-like conformations start to be approached at high diluton ($\nu \sim 0.8$). Unlike for the case $Wi=20$, the concentration has a strong effect on the conformations of the SCNPs at $Wi=200$. Concomitant and consistently with the shrinking found in the gyration radius (see data for $Wi=200$ in Figure~\ref{fig:poly-Rgmg-dens}a), the increase of the concentration above the overlap density leads to lower effective exponents $\nu \gtrsim 0.6$.

\begin{figure}[ht]
\centering
\includegraphics[width=0.42\linewidth]{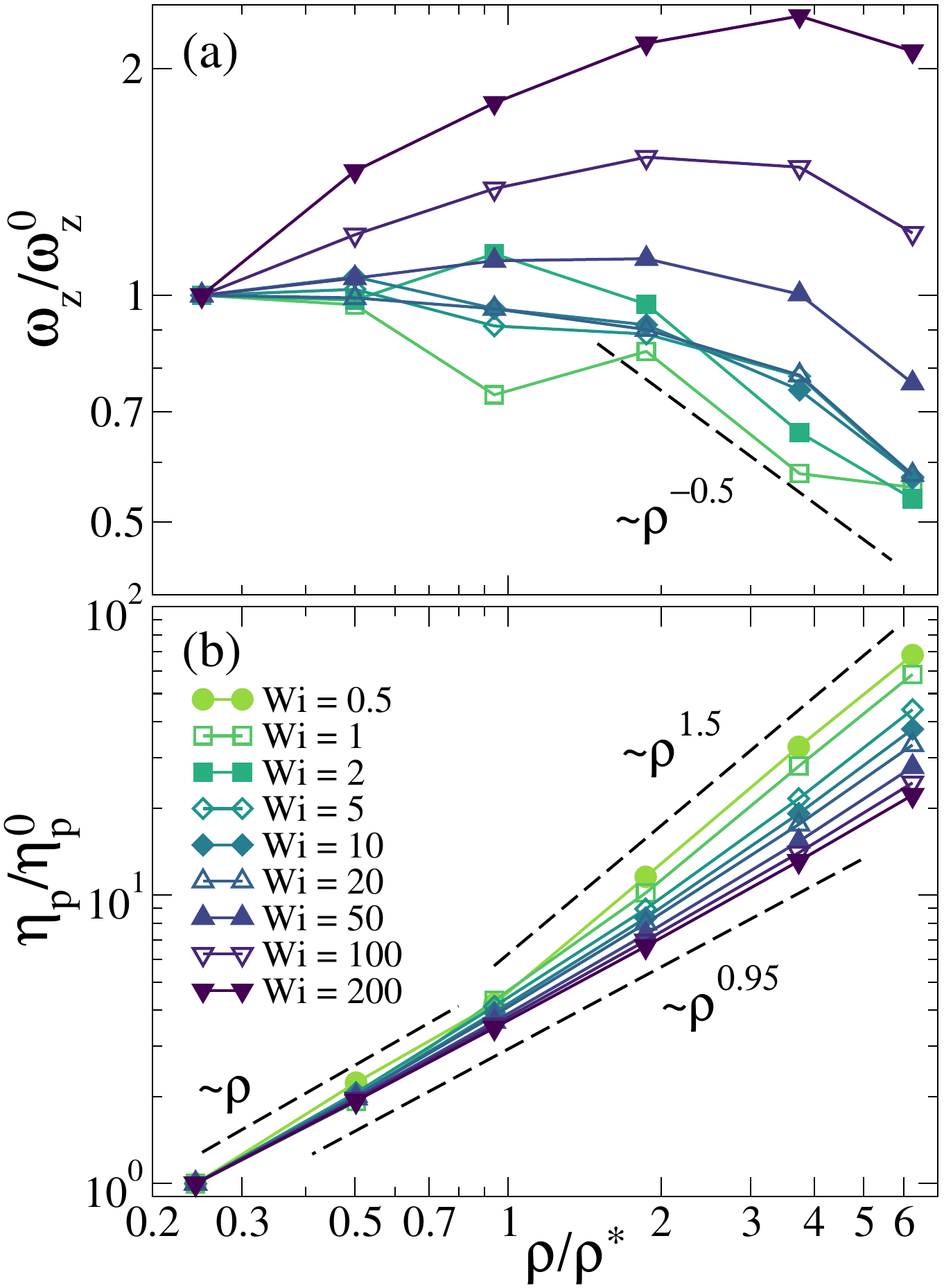}
\caption{As Figure~\ref{fig:poly-Rgmg-dens} for the rotational frecuency (a) and the
polymer contribution to the viscosity (b).}
\label{fig:poly-wzvisc-dens}
\end{figure}

Figure~\ref{fig:poly-wzvisc-dens} shows the rotational frequency and polymer contribution to the viscosity vs. the concentration
in the polydisperse solution.
Each data set corresponds to a fixed Weissenberg number, and data are normalized by the value ($\omega_z^0, \eta_{\rm p}^0$) at that $Wi$ and $\rho/\rho^{\ast} = 0.25$. 
The concentration dependence of $\omega_z$ at the different fixed values
of $Wi$ shows a good correlation with the $z$-component of the gyration tensor (Figure~S5c). Thus, swelling along the $z$-direction combined with the
concomitant shrinking in the $xy$-plane (Figure~S5a,b) seems to facilitate rotations of the SCNPs around the vorticity axis. Rotations are hindered when swelling and shrinking occur
along the $xy$-plane and $z$-direction, respectively.
As expected, the polymer contribution to the viscosity (Figure~\ref{fig:poly-wzvisc-dens}) is just proportional to the
concentration for $\rho \ll \rho^{\ast}$. At low and moderate values of $Wi$ it shows, around
the overlap density, a crossover from the linear to a power-law dependence, $\eta_{\rm p} \sim \rho^x$.
The exponent at $Wi \lesssim 1$  is $x=1.5$, which is intermediate between the values for linear
chains in equilibrium and semidilute solution at good ($x=1.3$) and $\theta$-solvent ($x=2$) 
conditions \cite{Rubinstein2003}. No significant crossover in the concentration dependence 
of the viscosity is found for the largest Weissenberg numbers $Wi \gtrsim 100$, for which a quasi-linear dependence $x=0.95$ is found.
Similar results for $\eta_p$ are found in the topologically monodisperse solutions (see Figure~S7 in the SI). 
The trends in the observed exponents can be rationalized by a rough scaling argument 
for unentangled semidilute solutions \cite{Rubinstein2003}. For macromolecular objects scaling
as $R \sim N^{\nu}$, with $R$ and $N$ their size and number of monomers respectively, their overlap concentration
should scale as $\rho^{\ast} \sim N R^{-3} \sim N^{1 -3\nu}$. Since above the overlap concentration
$\eta_{\rm p} \sim (\rho/\rho^{\ast})^x$,  we have $\eta_{\rm p} \sim \rho^x N^{(3\nu -1)x}$. On the other hand, in semidilute conditions the hydrodynamic interactions are screened beyond the mesh size, so that the viscosity should scale in a linear Rouse-like fashion \cite{Rubinstein2003}
with the macromolecular mass, $\eta_{\rm p} \sim N$.
Therefore the exponents $x$ and $\nu$ are related as $(3\nu -1)x =1$. According to this relation, the exponents for the viscosity
$x=1.5, 1.1, 0.95$ found at the representative values $Wi=1, 20, 200$ should originate from exponents for the molecular size
$\nu = 0.56, 0.64$ and 0.68, respectively. These are in good agreement with the analysis of $r(s)$, which gives $\nu =0.52$ for $Wi=1$ (not shown) and $\nu =0.63,0.66$ for $Wi=20$ and 200, respectively (Figure~\ref{fig:poly-rs}). 
Still, this agreement should be taken with caution due to the uncertainties in the determination of the $\nu$-values.

\section{III. Discussion}

The complex behavior of the concentration dependence of the size and viscosity of the SCNPs is inherent to their molecular architecture. 
Figure~S8 in the SI shows the corresponding results for semidilute of solutions of linear chains under shear
(data from Ref.~\citen{Huang2010mac}, see SI for details). 
As expected, in equilibrium ($Wi=0$) crowding leads to shrinking. However, once the Weissenberg number is sufficiently high ($Wi>2$)
the linear chains swell by increasing the concentration. No reentrance is observed and should not be found at higher $Wi$.
Though the swelling ratio $R_{\rm g}/R_{\rm g0}$ at  the highest investigated $Wi \approx 2400$ is lower than at moderate $Wi$'s,
it should be noted that at such a high $Wi$ swelling is just limitted by the fact that the chains at high dilution
are already close to rod-like objects and cannot be stretched much more. Increasing the concentration has a weaker effect on the viscosity of the sheared
solutions of linear chains (Figure~S8b) than in those made of SCNPs. According to the proposed relation $(3\nu -1)x =1$ (see above), the observed exponents
$x=1.2$ (low concentration, low $Wi$) and $0.8$ (high concentration, high $Wi$) correspond to values $\nu = 0.61$ and 0.75, respectively.
This is consistent with the limits of self-avoing random walk ($\nu_{\rm F}=0.59$) and rod ($\nu_{\rm R}=1$) that 
should be approached in the former regimes.

A tentative explanation for the very different trends observed for SCNPs and linear chains in
Figure~\ref{fig:poly-Rgmg-dens}a and S8a is as follows. When $Wi$ is high and they are strongly stretched, the macromolecules
respond to an increase of the concentration by stretching even more, because the `pseudonematic' ordering leads to a gain in vibrational (though side oscillations) and translational entropy that compensates the loss of conformational entropy induced by the stretching. This effect continues in the linear chains by further increasing the concentration, since the persistence of the quasi-rod conformations is not hindered by the neighboring
chains, and tumbling motions can be performed by sliding of one piece of the chain over the other without thickening significantly the
cross-section. This is not the case in the SCNPs, since beyond some point they cannot optimize packing through 
further increasing their elongation, which is impeded by the permanent cross-links in their architectures (25\% of cross-linking 
in the simulated systems). For this reason tumbling also involves adopting transient conformations that are much more compact than
those for linear chains (see right bottom panels in Figures~S3 and S4) and that coexist with the elongated ones, limiting the extension of the latter
and leading, in average, to smaller molecular sizes than at lower concentration. 
The presence of transient compact conformations across the solution and at all times is 
illustrated in Movies M1-M3 in the SI, which  show the dynamics of the monodisperse solution of SCNPs with middle asphericity, at $Wi=200$ and $\rho/\rho^{\ast}=3.74$.
The SCNPs are colored according to their instantaneous values of $R_{\rm g}$ as in Figures~S3 and S4. Movie M1 displays all the SCNPs in the solution.
Movie M2 shows, for the sake of clarity, only the SCNPs whose instantaneous position of the center-of-mass is within a fixed slice perpendicular 
to the $z$-axis and of width $\Delta z=10$. Movie M3 shows the trajectory of a selected SCNP. The big beads in M3 are the couple of, in average,
most distant mononomers in the SCNP, and are depicted in different colors to highlight tumbling motion.

\begin{figure}[ht]
\centering
\includegraphics[width=0.44\linewidth]{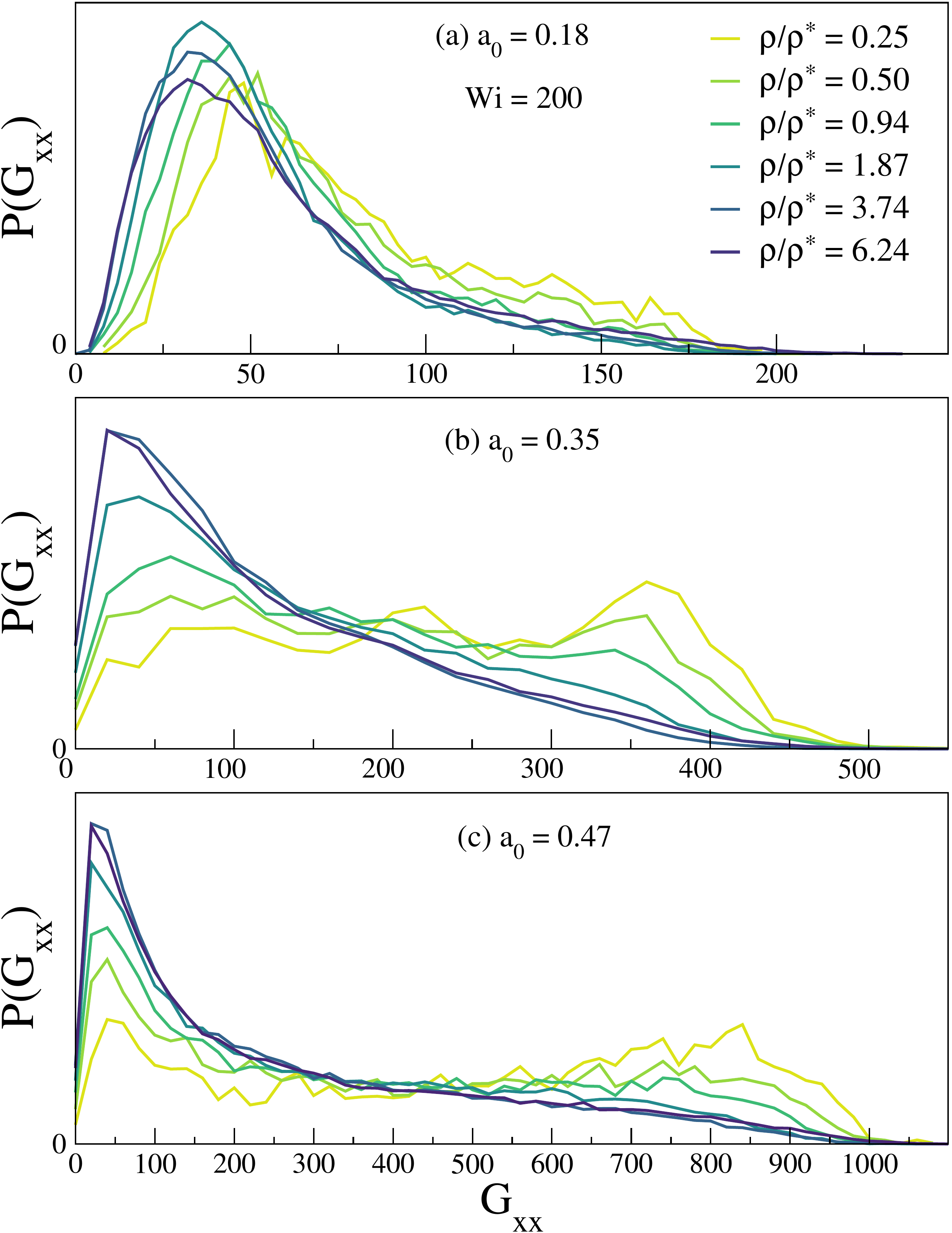}
\caption{Distribution of instantaneous $x$-components of the inertia tensor for the monodisperse
solutions, at high Weissenberg number $Wi=200$, of SCNPs with equilibrium asphericities $a_0 = 0.18$ (a), 0.34 (b) and 0.47 (c).
Each data set corresponds to a value of the concentration (see legend).}
\label{fig:mono-histogGxx}
\end{figure}

\begin{figure}[ht]
\centering
\includegraphics[width=0.44\linewidth]{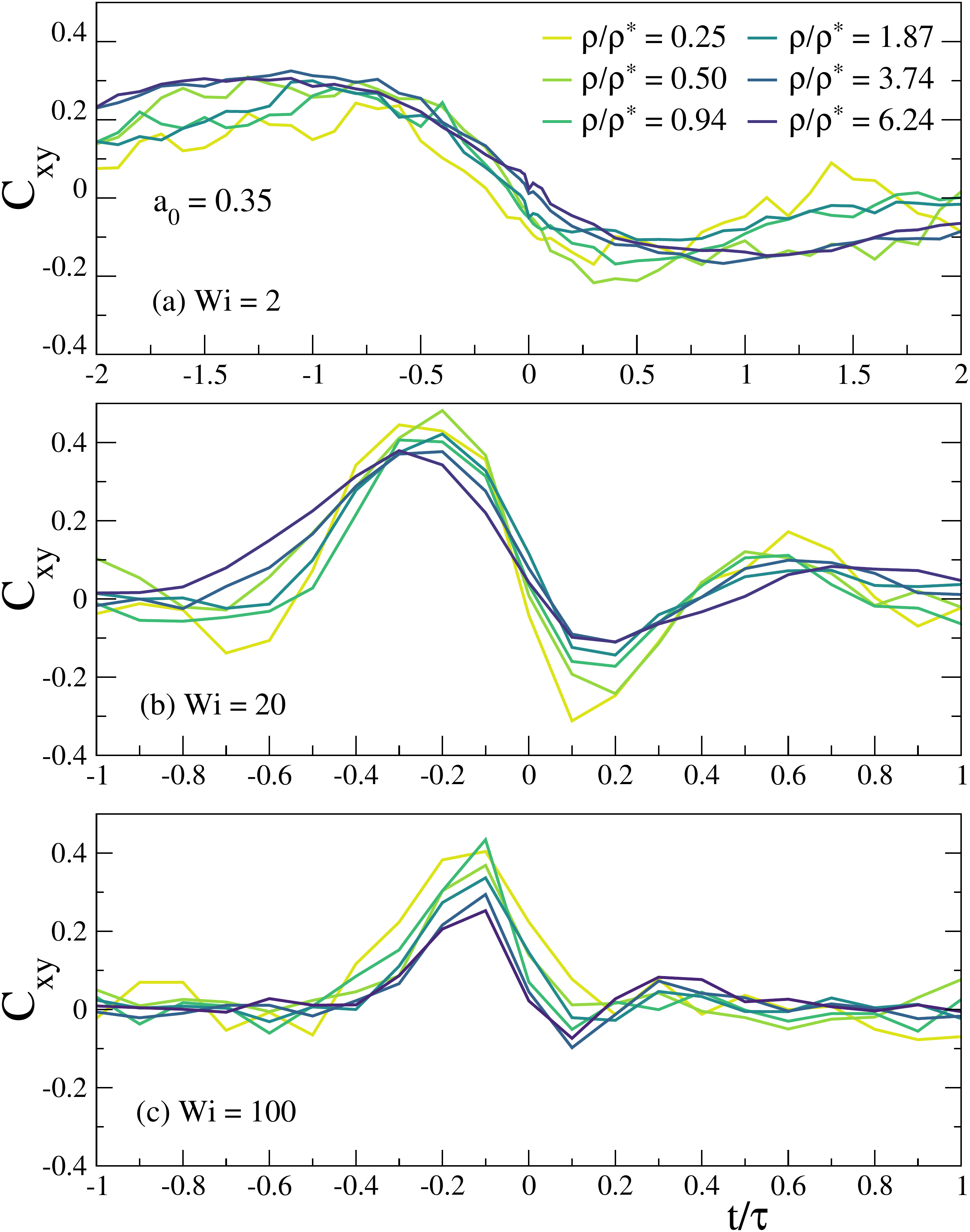}
\caption{Cross-correlator $C_{xy}(t)$ for the monodisperse solutions of SCNPs 
with middle asphericity $a_0=0.34$, at Weissenberg numbers $Wi=2, 20$ and 100 (panels (a), (b) and (c), respectively). 
Each data set corresponds to a fixed value of the concentration (see legend).}
\label{fig:mono-cxy}
\end{figure}

More insight about the reduction of the SCNP size by increasing the concentration at high $Wi$
can be obtained by analyzing the distribution of instantaneous configurations and characterizing the tumbling dynamics. 
Figure~\ref{fig:mono-histogGxx} shows the distribution of the instantenous
values of the $x$-component of the gyration tensor, $G_{\rm xx}$, at fixed $Wi=200$ in the monodisperse solutions of low, middle
and high asphericity. As can be seen, crowding leads to a higher presence of the least elongated conformations (low $G_{xx}$),
and in particular breaks the flat distribution (expected for well-defined tumbling motion)
found at low concentration for the sparse SCNPs (panels (b) and (c)).
Figure~\ref{fig:mono-cxy} shows the cross-correlator $C_{xy}$ of the $x$- and $y$-components of the gyration tensor for the monodisperse solutions with middle asphericity, 
at $Wi=2, 20$ and 100. The correlator is calculated as
\begin{equation}
C_{xy}(t) = \frac{\langle \delta G_{xx}(0) \delta G_{yy}(t) \rangle}{\sqrt{\langle\delta G_{xx}^2(0)\rangle \langle \delta G_{yy}^2(0)\rangle}} \, , 
\end{equation}
with  $\delta G_{\mu\mu} = G_{\mu\mu} - \langle G_{\mu\mu}\rangle$ the fluctuation of $G_{\mu\mu}$ around
its mean value $\langle G_{\mu\mu} \rangle$. The correlator $C_{xy}(t)$ is a useful observable for detecting tumbling dynamics in the motion of polymers under shear flow.
Tumbling is manifested as negative
anti-correlation peaks \cite{Huang2011,Chen2017}. These are found in Figure~\ref{fig:mono-cxy}, confirming the presence of tumbling. However, for high Weissenberg numbers the intensity of the peaks decays by increasing the concentration, showing that crowding has the effect of
reducing the contribution of tumbling to the motion of the SCNPs. 
This observation, together with that of Figure~\ref{fig:mono-histogGxx}, shows that transient compact conformations at high $Wi$ have a longer lifetime
when the solution becomes more crowded, hindering the stretching of the most elongated ones and leading to the mean shrinking of the SCNP size at high $Wi$ observed in Figure~\ref{fig:poly-Rgmg-dens} and S6. Still, at high $Wi$ the monotonic shrinking with increasing crowding stops above some concentration, and reentrant behavior is observed. The reason why the SCNPs start to swell at that point is not clear. 
It might be related with the microsegregation into domains
of low and high instantaneous values of $R_{\rm g}$ (Figures~S3 and S4). The development of these domains at  high values of both $Wi$ and $\rho$ may facilitate stretching and packing of the most elongated conformations, leading to the observed swelling.

\section{IV. Conclusions}

In summary, we have investigated for the first time the effect of the shear flow on the structural and dynamic properties of SCNPs in, 
semidilute, crowded solutions. We have characterized the dependence of several conformational and dynamic observables on the shear rate and the concentration.
The emerging physical scenario exhibits remarkable differences with those of topologically simple objects such as linear chains.
Whereas for the latter the shear-rate dependence is marginally dependent on the concentration, two clearly different scaling regimes are found for the SCNPs below and above the overlap concentration. Furthermore, crowding the solutions of SCNPs at fixed shear rate leads to a complex non-monotonic scenario
for the molecular size, in contrast to the case of linear chains, for which increasing the concentration at moderate or high shear rate always leads to swelling. The fractal globular conformations adopted by the SCNPs in equilibrium, originating from the topological interactions (non-concatenation constraint) that reduce interpenetrability in comparison with linear chains, have their counterpart in the strongly sheared solutions
as transient compact conformations, which hinder the stretching of the most elongated conformations. These compact conformations naturally arise from the cross-linked character of the SCNPs, which limit their maximum extension far below the rod-like limit,
and have a longer lifetime at high concentration due to partial supression of tumbling. 
This effect, together with the lower penetrability of the SCNPs arising from
the topological interactions, is at the origin of the rather different response to shear and crowding of solutions of SCNPs 
with respect to those of simple linear chains.

Beyond the consequences on the field of non-linear rheology of complex macromolecules, our system may have applications
as a simple model of IDPs under strong shear. Indeed IDPs should share more analogies with SCNPs under shear flow than in equilibrium: 
shear may break \cite{Bekard2011} the {\it ordered} domains of IDPs (this order being absent in SCNPs) that in equilibrium are stabilized through physical interactions (hydrogen bonds, electrostatic, assembly of hydrophobic groups, etc), 
whereas the chemical `cross-links' in the IDP structure (such as disulfide bonds) will remain.

\begin{acknowledgement}

We acknowledge support from the projects PGC2018-094548-B-I00 (MCIU/AEI/FEDER, UE) and IT-1175-19 (Basque Government, Spain).
We thank Arash Nikoubashman for useful discussions.

\end{acknowledgement}

\begin{suppinfo}
\begin{itemize}

\item Model and simulation details; SCNP architectures and simulation setup; snapshots of the simulation box
at several concentration and Weissenberg numbers; concentration dependence of molecular sizes and viscosities for SCNPs
and linear chains (PDF).

\item M1: Trajectory of the monodisperse solution of SCNPs with middle asphericity 
at $Wi=200$ and $\rho/\rho^{\ast}=3.74$ (MPG).

\item M2: As M1 for a slice perpendicular to the vorticity axis (MPG).

\item M3: As M1, highlighting the tumbling motion of a selected SCNP (MPG).

\end{itemize}
\end{suppinfo}

\newpage

\providecommand{\latin}[1]{#1}
\providecommand*\mcitethebibliography{\thebibliography}
\csname @ifundefined\endcsname{endmcitethebibliography}
  {\let\endmcitethebibliography\endthebibliography}{}

\newpage

\noindent{\bf SUPPORTING INFORMATION}\\
\\
\noindent{\bf Model and Simulation Details}\\
\\
The precursors are linear chains of $N=200$ beads (`monomers'). A fraction $f = N_r/N = 0.25$ of them are reactive monomers that will
form the cross-links. The $N_{\rm r}$ reactive groups are randomly distributed along the linear contour of the precursor, with the constraint that two
reactive groups are never placed consecutively in order to avoid trivial cross-links. 
The interactions are those of the bead-spring model \cite{Kremer1990bis}. Thus, the non-bonded interactions between any two given monomers 
(reactive or non-reactive) are given by a purely repulsive Lennard-Jones (LJ) potential,
\begin{equation}
U^{\rm LJ}(r) = 4\epsilon \left[ \left(\frac{\sigma}{r}\right)^{12} -\left(\frac{\sigma}{r}\right)^{6} +\frac{1}{4}\right]  .
\label{eq:vlj}
\end{equation}
Here $\epsilon/k_{\rm B}T = 1$ and $\sigma = 1$ set the units of energy and length, respectively. The potential is purely repulsive with no minimum
(mimicking pure excluded volume interactions in implicit good solvent) by using a cutoff distance $r_{\rm c} = 2^{1/6}\sigma$, which moreover guarantees
the continuity of the potential and forces at the cutoff. 

Bonded monomers along the contour of the chain and cross-linked monomers interact via a finitely 
extensible nonlinear elastic (FENE) potential \cite{Kremer1990bis}, 
\begin{equation}
U^{\rm FENE}(r) = - \epsilon K_{\rm F} R_0^2 \ln \left[ 1 - \left(  \frac{r}{R_0}\right)^2 \right] \, , 
\label{eq:fene}
\end{equation}
with $K_{\rm F} = 15\sigma^{-2}$ and $R_0 = 1.5\sigma$. This combination of LJ and FENE potentials limits the fluctuation of bonds and guarantees chain uncrossability \cite{Kremer1990bis}. 

We generate a set of 200 SCNPs via intramolecular cross-linking of 200 equilibrium realizations of the precursor. The cross-linking simulations
are performed in implicit solvent without hydrodynamic interactions, 
under Langevin dynamics following the scheme of Ref.~\citen{Izaguirre2001}. The precursors are coupled
to the same bath but do not interact with each other. Therefore cross-linking is purely intramolecular by construction, mimicking synthesis in the limit of high dilution. Cross-linking is irreversible and monovalent. Thus, a permanent bond between two reactive groups is formed when they are
separated by less than the capture distance $r_{\rm b} = 1.3\sigma$ and with the condition that none of them have already formed a bond with another
reactive group. A random selection is made when there are multiple candidates to form a bond within the capture distance. When the bond is formed
the involved monomers interact through the FENE potential for the remainder of the simulation. 
Figure S1 shows typical snapshots, in equilibrium at high dilution, of 6 topologically different SCNPs, covering the whole range of the obtained
distribution of asphericities \cite{Formanek2019bis}. The asphericity $0 \leq a_0 \leq 1$ of each SCNP is calculated at equilibrium (zero shear) and in the limit of high dilution (isolated SCNPs) as
\begin{equation}
a_0 = \frac{(\lambda_2-\lambda_1)^2 + (\lambda_3-\lambda_1)^2 + (\lambda_3-\lambda_2)^2}{2(\lambda_1+\lambda_2+\lambda_3)^2} \, ,
\label{eq:asph}
\end{equation}
where $\lambda_1 \geq \lambda_2 \geq \lambda_3$ are the eigenvalues of the gyration tensor. Higher values of $a_0$ correspond to larger deviations from
the spherical shape.

Steric crowding in the simulations of the solutions emerges by switching on the intermolecular interactions
(through the non-bonded LJ interactions between monomers of different SCNPs).
For these systems we perform hybrid mesoscale simulations by coupling standard molecular dynamics (MD) for the SCNPs with multi-particle collision dynamics (MPCD) \cite{Malevanets1999bis, Malevanets2000} for the solvent. The latter ensures the correct resolution of hydrodynamic interactions.
The solvent is formed by $N_s$ point-like particles of mass $m$. The MPCD algorithm consists of two alternating steps governing the dynamics of the solvent.
In the {\it streaming} step the solvent particles are propagated ballistically for a time $h$:
\begin{equation}
\vec{ r}_i (t+\Delta t) = \vec{ r}_i(t) + h \vec{ v}_i(t) \, ,
\label{eq:streaming}
\end{equation} 
with $\vec{ r}_i$ and $\vec{ v}_i$ the position and velocity of the solvent particle $i$.
In the {\it collision} step linear momentum is exchanged as follows. First, all the particles (solvent and monomers) 
are sorted into cubic cells of volume $a^3$ and subjected to a rotation around a random axis by an angle $\alpha$ with respect to the center-of-mass 
velocity of the cell $\vec{ v}_{\rm cm}$, i.e.
\begin{equation}
\vec{ v}_i (t + \Delta t) = \vec{ v}_{\rm cm}(t) + \mathbf{R}(\alpha)\left(\vec{ v}_i(t) - \vec{ v}_{\rm cm}(t)\right) \, ,
\label{eq:SR} 
\end{equation} 
with $\mathbf{R}(\alpha)$ the rotation matrix.
This conserves the total mass, linear momentum and energy of the system. To satisfy Galilean invariance, the cubic grid used to sort the particles has to be shifted randomly ($-a/2 < \Delta x, \Delta y, \Delta z < a/2$) in each of the 3 directions at each collision step\cite{Ihle2001, Ihle2003}.

Homogeneous shear is simulated by imposing a linear shear profile, $\langle v_x(y) \rangle= \dot{\gamma} y $, with Lees-Edwards boundary conditions\cite{Lees1972bis}. In the former expression $\dot{\gamma}$ is the shear rate, $v_x$ is the component of the velocity in the flow direction
and $y$ is the coordinate in the simulation box along the gradient direction. 
Figure S2 is a scheme of the simulation setup, showing the velocity profile and indicating the flow ($x$) and gradient ($y$) directions. 
The vorticity direction $(z)$ is perpendicular to the plane of the scheme.

\begin{figure}[h!]
\centering
\includegraphics[width=0.9\linewidth]{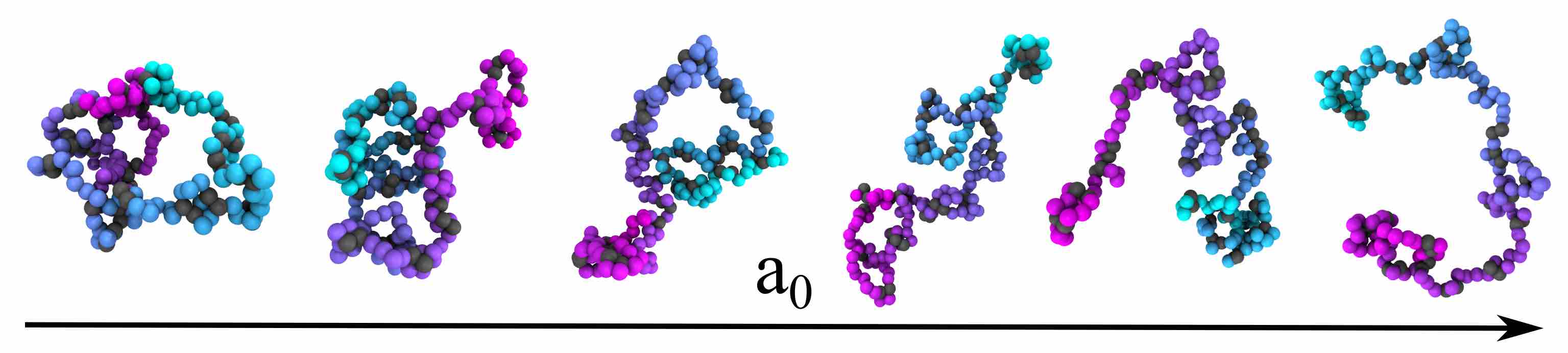}
\caption*{Figure~S1. Typical snapshots of SCNPs in equilibrium and high dilution,
with different values of the equilibrium asphericity $a_0$. From left to right, $a_0 = 0.17, 0.22, 0.34, 0.41, 0.47$ and 0.49.
Grey beads are cross-linked monomers. The rest of the monomers are colored, from magenta to cyan, according to their position
in the backbone of the linear precursor. }
\label{fgr:topol}
\end{figure}

\begin{figure}[h!]
\centering
\includegraphics[width=0.57\linewidth]{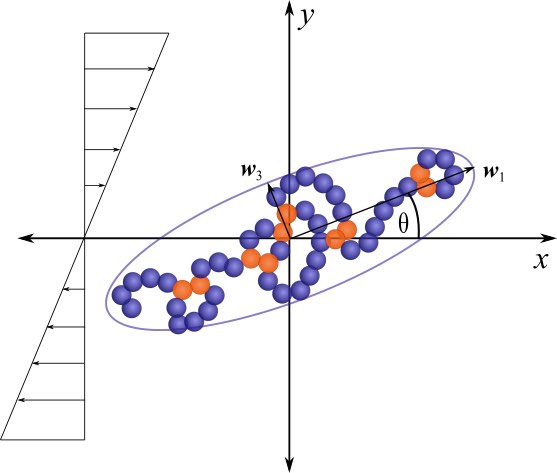}
\caption*{Figure~S2. Scheme of the simulation setup, indicating the fluid velocity profile and the eigenvalues and eigenvectors of the gyration tensor. The $x$-axis is the flow direction, $y$ is the gradient direction and $z$ -- perpendicular to the plane -- is the vorticity direction. $\theta$ is the angle between the largest eigenvector of the gyration tensor, $\vec{ w}_1$, and the direction of the flow. Reactive monomers forming cross-links are colored in orange, 
the rest are depicted in blue.}
\label{fgr:scheme}
\end{figure}

Since shear flow leads to viscous heating, a cell-level Maxwell-Boltzmann scaling thermostat is employed 
to keep the temperature of the fluid constant \cite{Mussawisade2005}.  
Between successive collision steps the SCNPs are propagated according to Newton equations of motion, which are integrated using the Velocity Verlet scheme \cite{Frenkel1996} with a time-step $\Delta t = 0.01$. The number of solvent particles per cell is $\rho = 5$, their mass is $m = 1$, while the mass of the solute monomers is $M = \rho m = 5$.  The remaining parameters are $\alpha =\SI{130}{\degree}$, $h = 0.1\sqrt{ma^2/k_BT}$ and $a = \sigma = 1$.
The former choices are standard values for MPCD and guarantee that the collisional viscosity dominates over the kinetic viscosity, and the hydrodynamic interactions are fully developed \cite{Mussawisade2005,Singh2014}.

We investigate the shear rates $\dot{\gamma} = 5 \times 10^{-4}, 10^{-3}, 2 \times 10^{-3}, 5 \times 10^{-3}, 10^{-2}$ and $2 \times 10^{-2}$. 
The reported Weissenberg numbers are defined as $Wi = \dot{\gamma}\tau$, where $\tau$ is the relaxation time at equilibrium and high dilution $\rho =0$. The value of $\tau$ is determined from the decay of the correlator of $R_{\rm g}$ \cite{Formanek2019bis}. We find 
$\tau \approx 10^4$ as the mean value of the polydisperse distribution, and  $\tau \approx 2\times 10^3, 10^4$ and $8\times 10^4$ for the SCNPs with, respectively, low ($a_0 = 0.18$), middle ($a_0 = 0.34$) and high ($a_0 = 0.47$) asphericity (values at $\dot{\gamma}=0$ and high dilution) \cite{Formanek2019bis} that we select for generating the topologically monodisperse solutions. 

The sides $L_x$ and $L_y = L_z$ of the simulation box 
are chosen so that in all cases $L_{\mu}$ is larger than $2\sqrt{G_{\mu\mu}}$, with $G_{\mu\mu}$ the diagonal component of the gyration
tensor in the $\mu$-direction and at the simulated shear rate. Thus, the box sides are varied in the ranges $28 \leq L_{y,z} \leq 56$ and $40 \leq L_x \leq 132$.
To generate the polydisperse solutions, SCNPs are taken from the previously created set and inserted 
in a large box at long mutual distances that prevent concatenation. For the monodisperse solutions, replicas of the same selected SCNP are
inserted in the box. After a short equilibration,
the box is very slowly compressed to the dimensions required by the selected shear rate and concentration. After equilibration
at $\dot{\gamma}=0$, the shear profile and Lees-Edwards boundary conditions are applied. When the studied observables (see the article) reach
steady states, accumulation runs are performed and the generated configurations are used in the analysis.
To improve statistics, several independent realizations of the box are simulated for each couple of values of the shear rate and concentration.
The number of independent runs is higher in the polydisperse systems, varying, for a fixed $\dot{\gamma}$, between 20 at the lowest concentration $\rho/\rho^{\ast} = 0.25$ and 5 at the highest one $\rho/\rho^{\ast} = 6.24$. The number of SCNPs in the simulation box varies, respectively, between
8 and the full set of 200. Different SCNPs are used in the independent runs of the polydisperse solution at low concentration, so that for every pair $(\dot{\gamma}, \rho)$ the topological distribution is correctly sampled.
\\
\\
\\
\\
\noindent{\bf Supplementary Figures}
\newpage

\begin{landscape}
\begin{figure}[h!]
\centering
\includegraphics[width=0.99\linewidth]{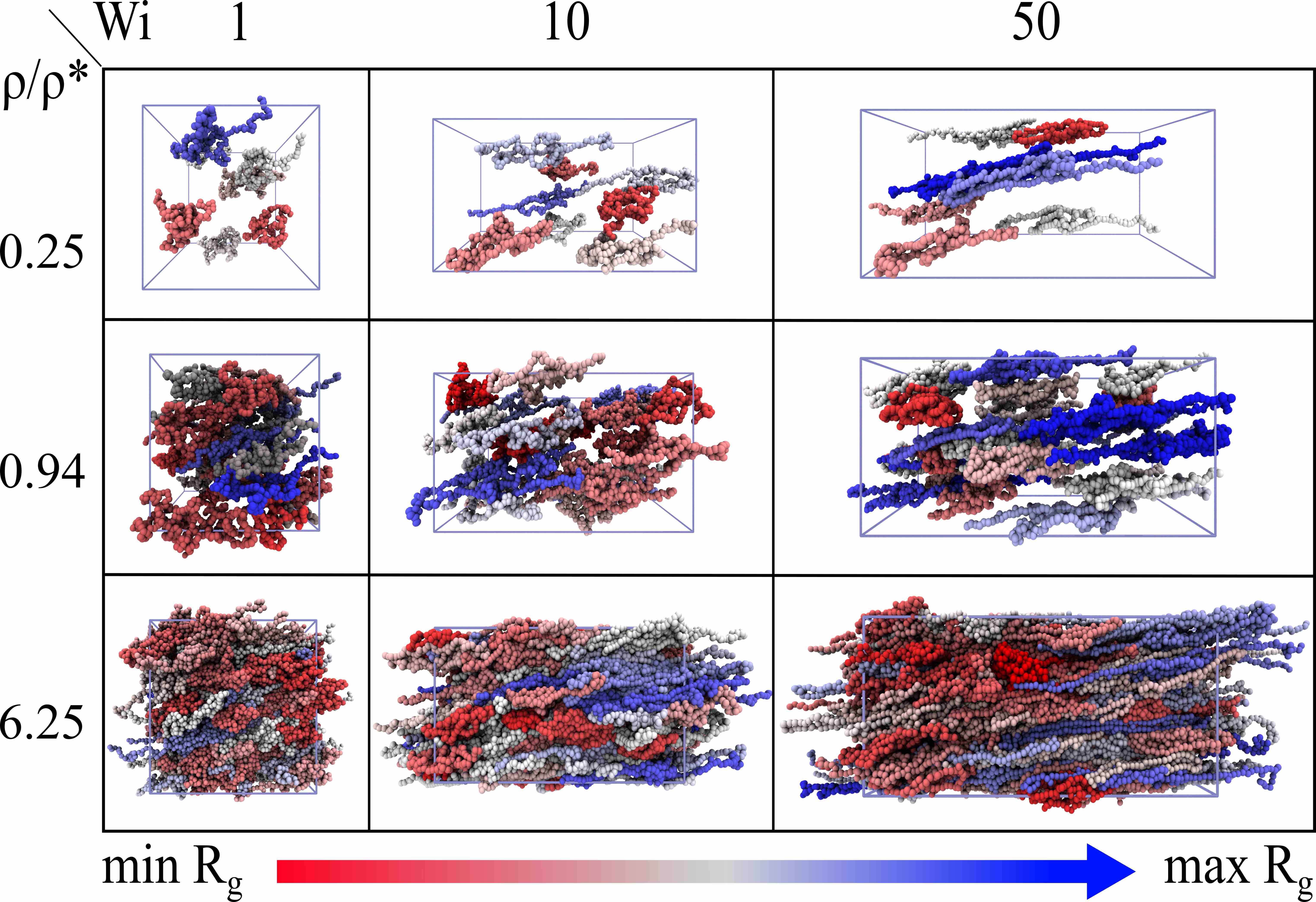}
\vspace{5 mm}
\caption*{Figure~S3. Snapshots of the simulation box for the monodisperse solution of SCNPs
with low equilibrium asphericity $a_0 = 0.18$, at different values of the Weissenberg number and the concentration.
All the SCNPs in the box are represented and they are colored according to their instantaneous
radius of gyration (dark red to dark blue from lower to higher $R_{\rm g}$, grey for medium size).}  
\label{fig:snap-asphlow}
\end{figure}
\end{landscape}

\newpage

\begin{landscape}
\begin{figure}[h!]
\centering
\includegraphics[width=0.99\linewidth]{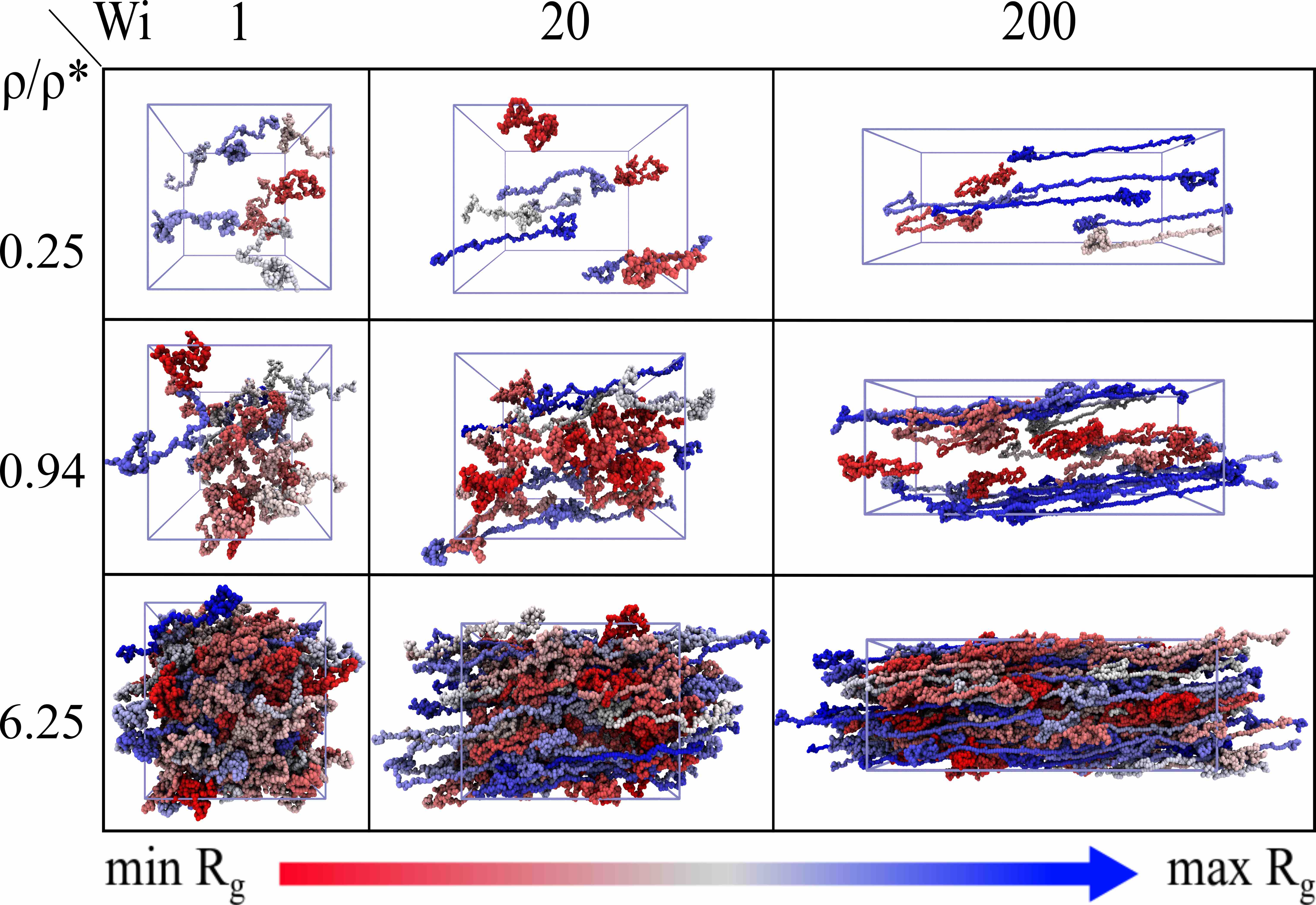}
\vspace{5 mm}
\caption*{Figure~S4. As Figure~S3 for the monodisperse solution of SCNPs with high equilibrium asphericity $a_0 = 0.47$.}  
\label{fig:snap-asphhigh}
\end{figure}
\end{landscape}

\newpage

\begin{figure}[h!]
\centering
\includegraphics[width=0.50\linewidth]{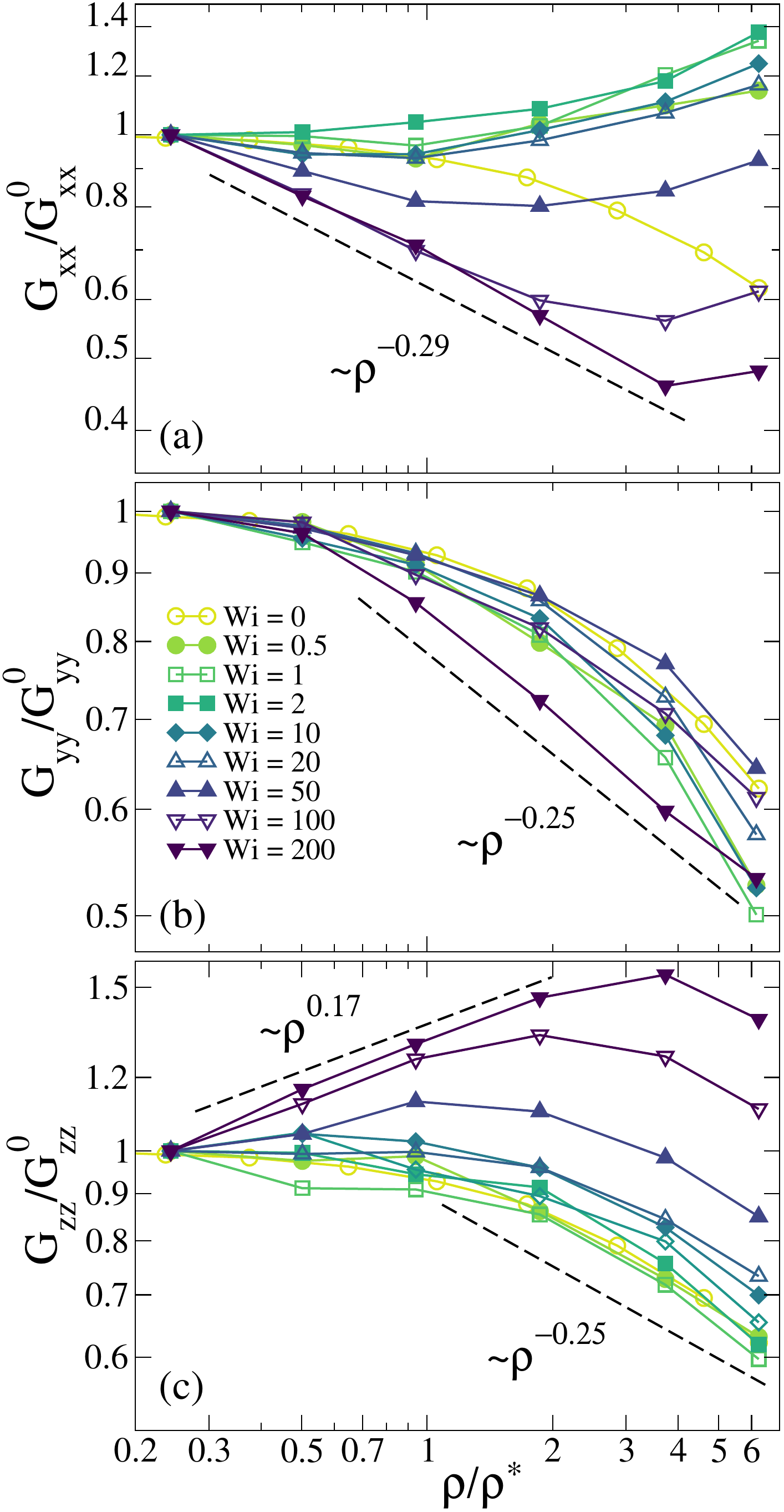}
\caption*{Figure~S5. For the SCNPs in the polydisperse solutions,
components of the gyration tensor vs. the concentration. 
Each data set corresponds to a fixed Weissenberg number (see legend) and is normalized by the
value ($G_{\mu\mu}^0$) at its corresponding $Wi$ and concentration $\rho/\rho^{\ast}=0.25$.
Dashed lines represent power laws.}
\label{fig:poly-Gallmag-weis}
\end{figure}

\newpage

\begin{figure}[h!]
\centering
\includegraphics[width=0.53\linewidth]{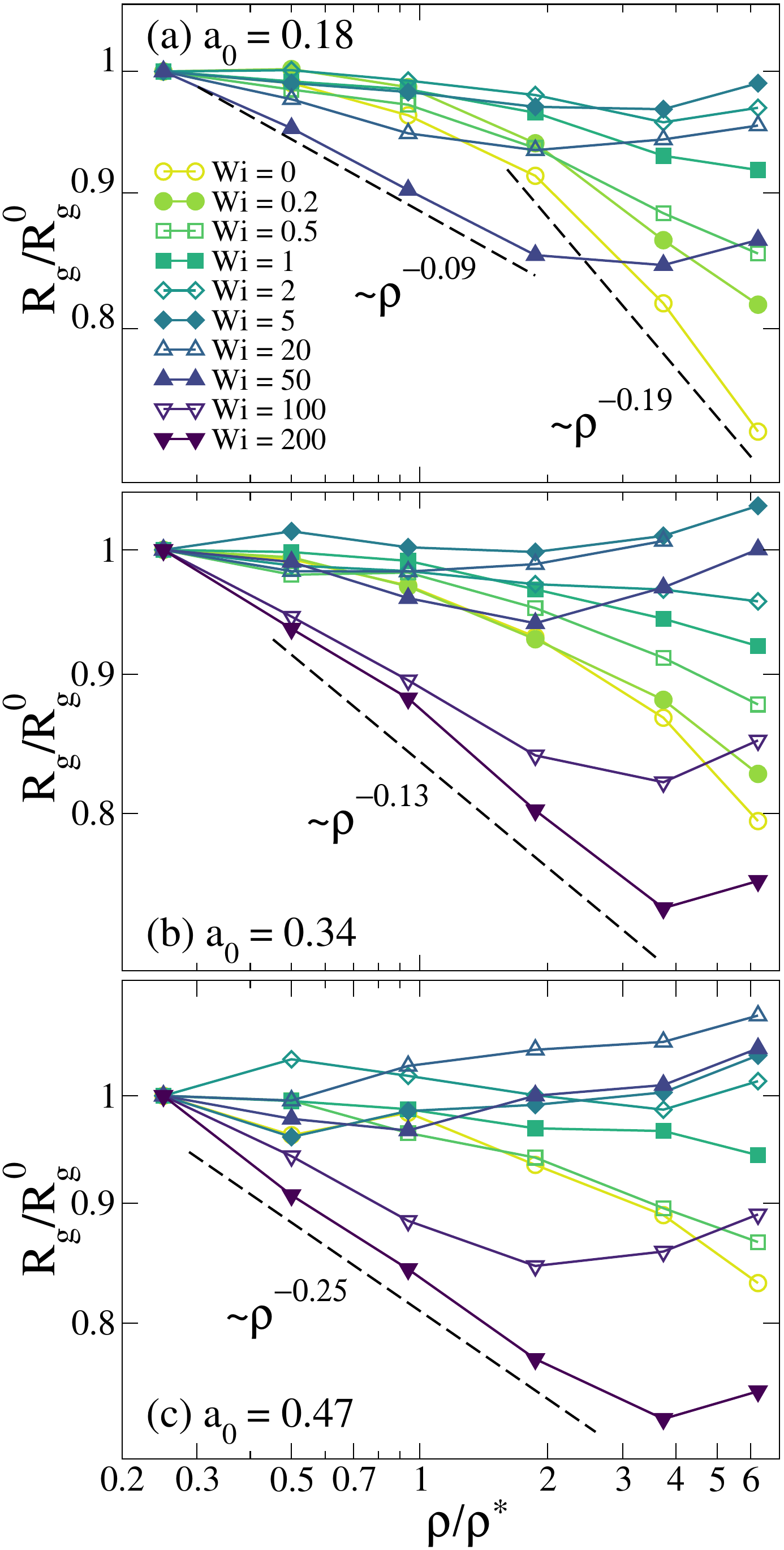}
\caption*{Figure~S6. For the SCNPs in the monodisperse solutions,
gyration radius vs. the concentration. Data are given for equilibrium asphericities
$a_0 = 0.18$ (a), 0.34 (b) and 0.47 (c).
Each data set corresponds to a fixed Weissenberg number (see legend) and is normalized by the
value ($R_{\rm g0}$) at its corresponding $Wi$ and concentration $\rho/\rho^{\ast}=0.25$.
Dashed lines represent power laws.} 
\label{fig:poly-wzvisc-weis}
\end{figure}

\newpage

\newpage

\begin{figure}[h!]
\centering
\includegraphics[width=0.5\linewidth]{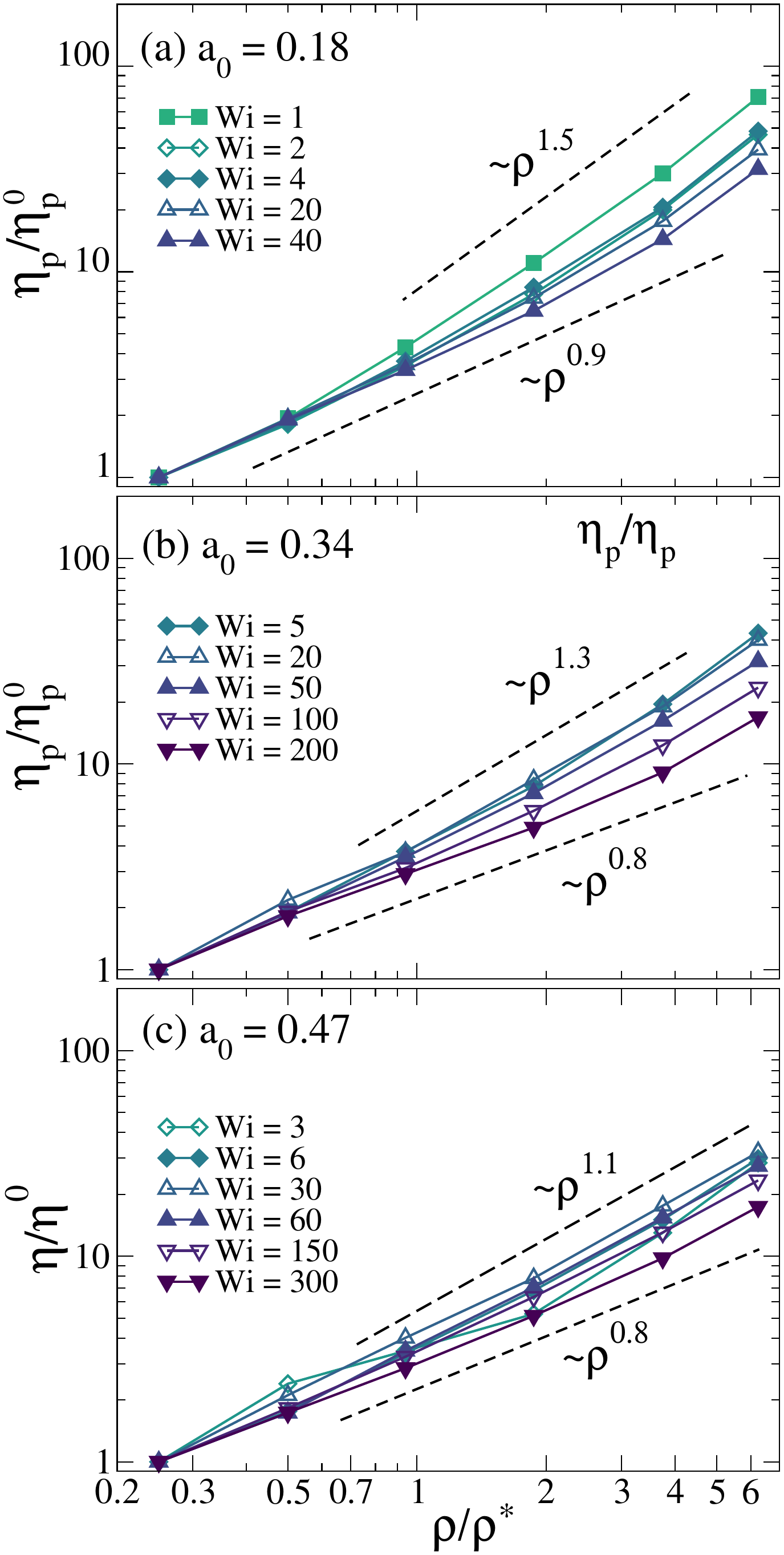}
\caption*{Figure~S7. Polymer contribution to the viscosity vs. the concentration in
monodisperse solutions of SCNPs with equilibrium asphericities $a_0 = 0.18$ (a), 0.34 (b) and 0.47 (c).
Each data set corresponds to a value of the Weissenberg number (see legends)
and is normalized by the value for that $Wi$ at the lowest concentration $\rho/\rho^{\ast}= 0.25$.}
\label{fig:mono-viscos-dens}
\end{figure}

\newpage

\begin{figure}[h!]
\centering
\includegraphics[width=0.5\linewidth]{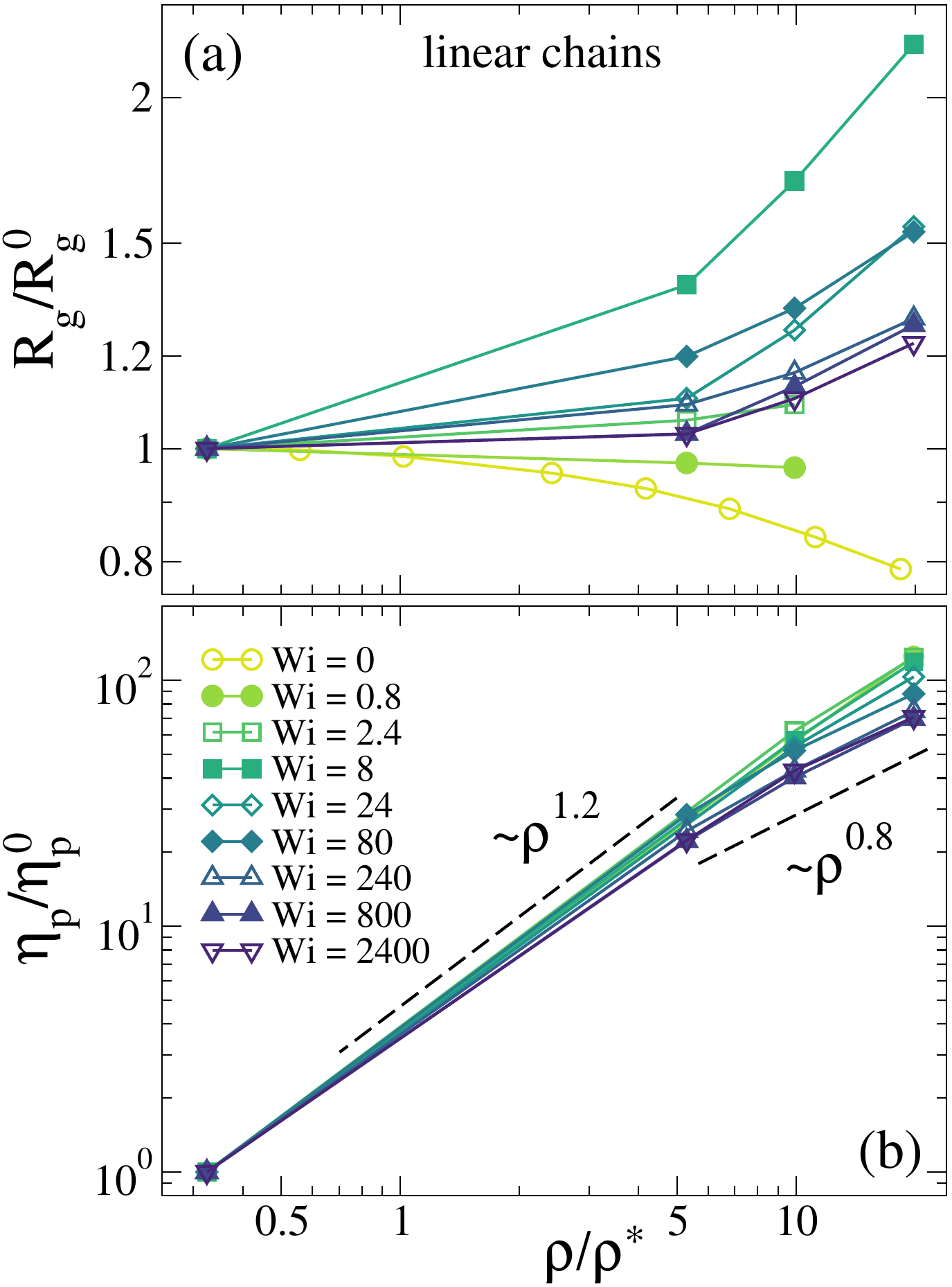}
\caption*{Figure~S8. Radius of gyration (a) and polymer contribution to the viscosity (b)
vs. the concentration, in solutions of linear chains (data are taken from Ref.~\citen{Huang2010macbis}).
Each data set corresponds to a fixed value of the Weissenberg number $Wi$ (see legends),
and is normalized by the value for that $Wi$ at the lowest 
concentration $\rho/\rho^{\ast}\approx 0.3$. It must be noted that our definition of the overlap concentration
is $\pi/6$ times the one used in  Ref.~\citen{Huang2010macbis}, so we have rescaled the data there by such a factor.}
\label{fig:linear-Rg-viscos-dens}
\end{figure}

\newpage
\providecommand{\latin}[1]{#1}
\providecommand*\mcitethebibliography{\thebibliography}
\csname @ifundefined\endcsname{endmcitethebibliography}
  {\let\endmcitethebibliography\endthebibliography}{}


\begin{mcitethebibliography}{56}
\providecommand*\natexlab[1]{#1}
\providecommand*\mciteSetBstSublistMode[1]{}
\providecommand*\mciteSetBstMaxWidthForm[2]{}
\providecommand*\mciteBstWouldAddEndPuncttrue
  {\def\EndOfBibitem{\unskip.}}
\providecommand*\mciteBstWouldAddEndPunctfalse
  {\let\EndOfBibitem\relax}
\providecommand*\mciteSetBstMidEndSepPunct[3]{}
\providecommand*\mciteSetBstSublistLabelBeginEnd[3]{}
\providecommand*\EndOfBibitem{}
\mciteSetBstSublistMode{f}
\mciteSetBstMaxWidthForm{subitem}{(\alph{mcitesubitemcount})}
\mciteSetBstSublistLabelBeginEnd
  {\mcitemaxwidthsubitemform\space}
  {\relax}
  {\relax}

\bibitem[Pomposo(2017)]{PomposoSCNPbook}
Pomposo,~J.~A., Ed. \emph{Single-Chain Polymer Nanoparticles: Synthesis,
  Characterization, Simulations, and Applications}; John Wiley \& Sons:
  Weinheim, Germany, 2017\relax
\mciteBstWouldAddEndPuncttrue
\mciteSetBstMidEndSepPunct{\mcitedefaultmidpunct}
{\mcitedefaultendpunct}{\mcitedefaultseppunct}\relax
\EndOfBibitem
\bibitem[Terashima \latin{et~al.}(2011)Terashima, Mes, De~Greef, Gillissen,
  Besenius, Palmans, and Meijer]{terashima2011}
Terashima,~T.; Mes,~T.; De~Greef,~T. F.~A.; Gillissen,~M. A.~J.; Besenius,~P.;
  Palmans,~A. R.~A.; Meijer,~E.~W. Single-Chain Folding of Polymers for
  Catalytic Systems in Water. \emph{J. Am. Chem. Soc.} \textbf{2011},
  \emph{133}, 4742--4745\relax
\mciteBstWouldAddEndPuncttrue
\mciteSetBstMidEndSepPunct{\mcitedefaultmidpunct}
{\mcitedefaultendpunct}{\mcitedefaultseppunct}\relax
\EndOfBibitem
\bibitem[Perez-Baena \latin{et~al.}(2013)Perez-Baena, Barroso-Bujans, Gasser,
  Arbe, Moreno, Colmenero, and Pomposo]{perez2013endowing}
Perez-Baena,~I.; Barroso-Bujans,~F.; Gasser,~U.; Arbe,~A.; Moreno,~A.~J.;
  Colmenero,~J.; Pomposo,~J.~A. Endowing Single-Chain Polymer Nanoparticles
  with Enzyme-Mimetic Activity. \emph{ACS Macro Lett.} \textbf{2013}, \emph{2},
  775--779\relax
\mciteBstWouldAddEndPuncttrue
\mciteSetBstMidEndSepPunct{\mcitedefaultmidpunct}
{\mcitedefaultendpunct}{\mcitedefaultseppunct}\relax
\EndOfBibitem
\bibitem[Huerta \latin{et~al.}(2013)Huerta, Stals, Meijer, and
  Palmans]{huerta2013}
Huerta,~E.; Stals,~P. J.~M.; Meijer,~E.~W.; Palmans,~A. R.~A. Consequences of
  Folding a Water-Soluble Polymer Around an Organocatalyst. \emph{Angew. Chem.
  Int. Ed.} \textbf{2013}, \emph{52}, 2906--2910\relax
\mciteBstWouldAddEndPuncttrue
\mciteSetBstMidEndSepPunct{\mcitedefaultmidpunct}
{\mcitedefaultendpunct}{\mcitedefaultseppunct}\relax
\EndOfBibitem
\bibitem[Tooley \latin{et~al.}(2015)Tooley, Pazicni, and Berda]{tooley2015}
Tooley,~C.~A.; Pazicni,~S.; Berda,~E.~B. Toward a tunable synthetic [FeFe]
  hydrogenase mimic: single-chain nanoparticles functionalized with a single
  diiron cluster. \emph{Polym. Chem.} \textbf{2015}, \emph{6}, 7646--7651\relax
\mciteBstWouldAddEndPuncttrue
\mciteSetBstMidEndSepPunct{\mcitedefaultmidpunct}
{\mcitedefaultendpunct}{\mcitedefaultseppunct}\relax
\EndOfBibitem
\bibitem[Hamilton and Harth(2009)Hamilton, and Harth]{hamilton2009}
Hamilton,~S.~K.; Harth,~E. Molecular dendritic transporter nanoparticle vectors
  provide efficient intracellular delivery of peptides. \emph{ACS Nano}
  \textbf{2009}, \emph{3}, 402--410\relax
\mciteBstWouldAddEndPuncttrue
\mciteSetBstMidEndSepPunct{\mcitedefaultmidpunct}
{\mcitedefaultendpunct}{\mcitedefaultseppunct}\relax
\EndOfBibitem
\bibitem[Sanchez-Sanchez \latin{et~al.}(2013)Sanchez-Sanchez, Akbari, Moreno,
  Lo~Verso, Arbe, Colmenero, and Pomposo]{sanchez2013design}
Sanchez-Sanchez,~A.; Akbari,~S.; Moreno,~A.~J.; Lo~Verso,~F.; Arbe,~A.;
  Colmenero,~J.; Pomposo,~J.~A. Design and Preparation of Single-Chain
  Nanocarriers Mimicking Disordered Proteins for Combined Delivery of Dermal
  Bioactive Cargos. \emph{Macromol. Rapid Commun.} \textbf{2013}, \emph{34},
  1681--1686\relax
\mciteBstWouldAddEndPuncttrue
\mciteSetBstMidEndSepPunct{\mcitedefaultmidpunct}
{\mcitedefaultendpunct}{\mcitedefaultseppunct}\relax
\EndOfBibitem
\bibitem[Gillissen \latin{et~al.}(2012)Gillissen, Voets, Meijer, and
  Palmans]{gillisen2012single}
Gillissen,~M. A.~J.; Voets,~I.~K.; Meijer,~E.~W.; Palmans,~A. R.~A. Single
  chain polymeric nanoparticles as compartmentalised sensors for metal ions.
  \emph{Polym. Chem.} \textbf{2012}, \emph{3}, 3166--3174\relax
\mciteBstWouldAddEndPuncttrue
\mciteSetBstMidEndSepPunct{\mcitedefaultmidpunct}
{\mcitedefaultendpunct}{\mcitedefaultseppunct}\relax
\EndOfBibitem
\bibitem[Mackay \latin{et~al.}(2003)Mackay, Dao, Tuteja, Ho, Horn, Kim, and
  Hawker]{Mackay2003}
Mackay,~M.~E.; Dao,~T.~T.; Tuteja,~A.; Ho,~D.~L.; Horn,~B.~V.; Kim,~H.-C.;
  Hawker,~C.~J. Nanoscale effects leading to non-Einstein-like decrease in
  viscosity. \emph{Nat. Mater.} \textbf{2003}, \emph{2}, 762--766\relax
\mciteBstWouldAddEndPuncttrue
\mciteSetBstMidEndSepPunct{\mcitedefaultmidpunct}
{\mcitedefaultendpunct}{\mcitedefaultseppunct}\relax
\EndOfBibitem
\bibitem[Ba{\v{c}}ov{\'{a}} \latin{et~al.}(2017)Ba{\v{c}}ov{\'{a}}, Lo~Verso,
  Arbe, Colmenero, Pomposo, and Moreno]{Bacova2017}
Ba{\v{c}}ov{\'{a}},~P.; Lo~Verso,~F.; Arbe,~A.; Colmenero,~J.; Pomposo,~J.~A.;
  Moreno,~A.~J. The Role of the Topological Constraints in the Chain Dynamics
  in All-Polymer Nanocomposites. \emph{Macromolecules} \textbf{2017},
  \emph{50}, 1719--1731\relax
\mciteBstWouldAddEndPuncttrue
\mciteSetBstMidEndSepPunct{\mcitedefaultmidpunct}
{\mcitedefaultendpunct}{\mcitedefaultseppunct}\relax
\EndOfBibitem
\bibitem[Moreno \latin{et~al.}(2013)Moreno, Lo~Verso, Sanchez-Sanchez, Arbe,
  Colmenero, and Pomposo]{Moreno2013}
Moreno,~A.~J.; Lo~Verso,~F.; Sanchez-Sanchez,~A.; Arbe,~A.; Colmenero,~J.;
  Pomposo,~J.~A. Advantages of Orthogonal Folding of Single Polymer Chains to
  Soft Nanoparticles. \emph{{M}acromolecules} \textbf{2013}, \emph{46},
  9748--9759\relax
\mciteBstWouldAddEndPuncttrue
\mciteSetBstMidEndSepPunct{\mcitedefaultmidpunct}
{\mcitedefaultendpunct}{\mcitedefaultseppunct}\relax
\EndOfBibitem
\bibitem[Basasoro \latin{et~al.}(2016)Basasoro, Gonzalez-Burgos, Moreno, Verso,
  Arbe, Colmenero, and Pomposo]{Basasoro2016}
Basasoro,~S.; Gonzalez-Burgos,~M.; Moreno,~A.~J.; Verso,~F.~L.; Arbe,~A.;
  Colmenero,~J.; Pomposo,~J.~A. A Solvent-Based Strategy for Tuning the
  Internal Structure of Metallo-Folded Single-Chain Nanoparticles.
  \emph{Macromol. Rapid Commun.} \textbf{2016}, \emph{37}, 1060--1065\relax
\mciteBstWouldAddEndPuncttrue
\mciteSetBstMidEndSepPunct{\mcitedefaultmidpunct}
{\mcitedefaultendpunct}{\mcitedefaultseppunct}\relax
\EndOfBibitem
\bibitem[Arbe \latin{et~al.}(2016)Arbe, Pomposo, Moreno, LoVerso,
  Gonzalez-Burgos, Asenjo-Sanz, Iturrospe, Radulescu, Ivanova, and
  Colmenero]{Arbe2016}
Arbe,~A.; Pomposo,~J.; Moreno,~A.; LoVerso,~F.; Gonzalez-Burgos,~M.;
  Asenjo-Sanz,~I.; Iturrospe,~A.; Radulescu,~A.; Ivanova,~O.; Colmenero,~J.
  Structure and dynamics of single-chain nano-particles in solution.
  \emph{Polymer} \textbf{2016}, \emph{105}, 532 -- 544\relax
\mciteBstWouldAddEndPuncttrue
\mciteSetBstMidEndSepPunct{\mcitedefaultmidpunct}
{\mcitedefaultendpunct}{\mcitedefaultseppunct}\relax
\EndOfBibitem
\bibitem[Gonzalez-Burgos \latin{et~al.}(2018)Gonzalez-Burgos, Arbe, Moreno,
  Pomposo, Radulescu, and Colmenero]{GonzalezBurgos2018}
Gonzalez-Burgos,~M.; Arbe,~A.; Moreno,~A.~J.; Pomposo,~J.~A.; Radulescu,~A.;
  Colmenero,~J. Crowding the Environment of Single-Chain Nanoparticles: A
  Combined Study by SANS and Simulations. \emph{Macromolecules} \textbf{2018},
  \emph{51}, 1573--1585\relax
\mciteBstWouldAddEndPuncttrue
\mciteSetBstMidEndSepPunct{\mcitedefaultmidpunct}
{\mcitedefaultendpunct}{\mcitedefaultseppunct}\relax
\EndOfBibitem
\bibitem[Pomposo \latin{et~al.}(2014)Pomposo, Perez-Baena, Lo~Verso, Moreno,
  Arbe, and Colmenero]{Pomposo2014a}
Pomposo,~J.~A.; Perez-Baena,~I.; Lo~Verso,~F.; Moreno,~A.~J.; Arbe,~A.;
  Colmenero,~J. How Far Are Single-Chain Polymer Nanoparticles in Solution from
  the Globular State? \emph{ACS Macro Lett.} \textbf{2014}, \emph{3},
  767--772\relax
\mciteBstWouldAddEndPuncttrue
\mciteSetBstMidEndSepPunct{\mcitedefaultmidpunct}
{\mcitedefaultendpunct}{\mcitedefaultseppunct}\relax
\EndOfBibitem
\bibitem[Pomposo \latin{et~al.}(2017)Pomposo, Rubio-Cervilla, Moreno, Lo~Verso,
  Bacova, Arbe, and Colmenero]{Pomposo2017rever}
Pomposo,~J.~A.; Rubio-Cervilla,~J.; Moreno,~A.~J.; Lo~Verso,~F.; Bacova,~P.;
  Arbe,~A.; Colmenero,~J. Folding Single Chains to Single-Chain Nanoparticles
  via Reversible Interactions: What Size Reduction Can One Expect?
  \emph{Macromolecules} \textbf{2017}, \emph{50}, 1732--1739\relax
\mciteBstWouldAddEndPuncttrue
\mciteSetBstMidEndSepPunct{\mcitedefaultmidpunct}
{\mcitedefaultendpunct}{\mcitedefaultseppunct}\relax
\EndOfBibitem
\bibitem[Rubinstein and Colby(2003)Rubinstein, and Colby]{Rubinstein2003}
Rubinstein,~M.; Colby,~R.~H. \emph{{P}olymer {P}hysics}; {O}xford {U}niversity
  {P}ress: {O}xford, {U}.{K}., 2003; Vol.~23\relax
\mciteBstWouldAddEndPuncttrue
\mciteSetBstMidEndSepPunct{\mcitedefaultmidpunct}
{\mcitedefaultendpunct}{\mcitedefaultseppunct}\relax
\EndOfBibitem
\bibitem[Lo~Verso \latin{et~al.}(2014)Lo~Verso, Pomposo, Colmenero, and
  Moreno]{LoVerso2014}
Lo~Verso,~F.; Pomposo,~J.~A.; Colmenero,~J.; Moreno,~A.~J. Multi-orthogonal
  folding of single polymer chains into soft nanoparticles. \emph{{S}oft
  {M}atter} \textbf{2014}, \emph{10}, 4813--4821\relax
\mciteBstWouldAddEndPuncttrue
\mciteSetBstMidEndSepPunct{\mcitedefaultmidpunct}
{\mcitedefaultendpunct}{\mcitedefaultseppunct}\relax
\EndOfBibitem
\bibitem[Rabbel \latin{et~al.}(2017)Rabbel, Breier, and Sommer]{Rabbel2017}
Rabbel,~H.; Breier,~P.; Sommer,~J.-U. Swelling Behavior of Single-Chain Polymer
  Nanoparticles: Theory and Simulation. \emph{Macromolecules} \textbf{2017},
  \emph{50}, 7410--7418\relax
\mciteBstWouldAddEndPuncttrue
\mciteSetBstMidEndSepPunct{\mcitedefaultmidpunct}
{\mcitedefaultendpunct}{\mcitedefaultseppunct}\relax
\EndOfBibitem
\bibitem[Formanek and Moreno(2017)Formanek, and Moreno]{Formanek2017}
Formanek,~M.; Moreno,~A.~J. Effects of precursor topology and synthesis under
  crowding conditions on the structure of single-chain polymer nanoparticles.
  \emph{Soft Matter} \textbf{2017}, \emph{13}, 6430--6438\relax
\mciteBstWouldAddEndPuncttrue
\mciteSetBstMidEndSepPunct{\mcitedefaultmidpunct}
{\mcitedefaultendpunct}{\mcitedefaultseppunct}\relax
\EndOfBibitem
\bibitem[Oyarzun and Mognetti(2018)Oyarzun, and Mognetti]{Oyarzun2018}
Oyarzun,~B.; Mognetti,~B.~M. Efficient sampling of reversible cross-linking
  polymers: Self-assembly of single-chain polymeric nanoparticles. \emph{J.
  Chem. Phys.} \textbf{2018}, \emph{148}, 114110\relax
\mciteBstWouldAddEndPuncttrue
\mciteSetBstMidEndSepPunct{\mcitedefaultmidpunct}
{\mcitedefaultendpunct}{\mcitedefaultseppunct}\relax
\EndOfBibitem
\bibitem[Moreno \latin{et~al.}(2018)Moreno, Bacova, Verso, Arbe, Colmenero, and
  Pomposo]{Moreno2018}
Moreno,~A.~J.; Bacova,~P.; Verso,~F.~L.; Arbe,~A.; Colmenero,~J.;
  Pomposo,~J.~A. Effect of chain stiffness on the structure of single-chain
  polymer nanoparticles. \emph{J. Phys.: Condens. Matter} \textbf{2018},
  \emph{30}, 034001\relax
\mciteBstWouldAddEndPuncttrue
\mciteSetBstMidEndSepPunct{\mcitedefaultmidpunct}
{\mcitedefaultendpunct}{\mcitedefaultseppunct}\relax
\EndOfBibitem
\bibitem[Moreno \latin{et~al.}(2016)Moreno, Lo~Verso, Arbe, Pomposo, and
  Colmenero]{Moreno2016JPCL}
Moreno,~A.~J.; Lo~Verso,~F.; Arbe,~A.; Pomposo,~J.~A.; Colmenero,~J.
  Concentrated Solutions of Single-Chain Nanoparticles: A Simple Model for
  Intrinsically Disordered Proteins under Crowding Conditions. \emph{{J}.
  {P}hys. {C}hem. {L}ett.} \textbf{2016}, \emph{7}, 838--844\relax
\mciteBstWouldAddEndPuncttrue
\mciteSetBstMidEndSepPunct{\mcitedefaultmidpunct}
{\mcitedefaultendpunct}{\mcitedefaultseppunct}\relax
\EndOfBibitem
\bibitem[Oberdisse \latin{et~al.}(2019)Oberdisse, González-Burgos, Mendia,
  Arbe, Moreno, Pomposo, Radulescu, and Colmenero]{Oberdisse2019}
Oberdisse,~J.; González-Burgos,~M.; Mendia,~A.; Arbe,~A.; Moreno,~A.~J.;
  Pomposo,~J.~A.; Radulescu,~A.; Colmenero,~J. Effect of Molecular Crowding on
  Conformation and Interactions of Single-Chain Nanoparticles.
  \emph{Macromolecules} \textbf{2019}, \emph{52}, 4295--4305\relax
\mciteBstWouldAddEndPuncttrue
\mciteSetBstMidEndSepPunct{\mcitedefaultmidpunct}
{\mcitedefaultendpunct}{\mcitedefaultseppunct}\relax
\EndOfBibitem
\bibitem[Grosberg \latin{et~al.}(1988)Grosberg, Nechaev, and
  Shakhnovich]{Grosberg1988}
Grosberg,~A.~Y.; Nechaev,~S.~K.; Shakhnovich,~E.~I. The role of topological
  constraints in the kinetics of collapse of macromolecules. \emph{J. Phys.
  (Paris)} \textbf{1988}, \emph{49}, 2095--2100\relax
\mciteBstWouldAddEndPuncttrue
\mciteSetBstMidEndSepPunct{\mcitedefaultmidpunct}
{\mcitedefaultendpunct}{\mcitedefaultseppunct}\relax
\EndOfBibitem
\bibitem[Mirny(2011)]{Mirny2011}
Mirny,~L.~A. The fractal globule as a model of chromatin architecture in the
  cell. \emph{Chromosome Res.} \textbf{2011}, \emph{19}, 37--51\relax
\mciteBstWouldAddEndPuncttrue
\mciteSetBstMidEndSepPunct{\mcitedefaultmidpunct}
{\mcitedefaultendpunct}{\mcitedefaultseppunct}\relax
\EndOfBibitem
\bibitem[Halverson \latin{et~al.}(2014)Halverson, Smrek, Kremer, and
  Grosberg]{Halverson2014}
Halverson,~J.~D.; Smrek,~J.; Kremer,~K.; Grosberg,~A.~Y. From a melt of rings
  to chromosome territories: the role of topological constraints in genome
  folding. \emph{Rep. Prog. Phys.} \textbf{2014}, \emph{77}, 022601\relax
\mciteBstWouldAddEndPuncttrue
\mciteSetBstMidEndSepPunct{\mcitedefaultmidpunct}
{\mcitedefaultendpunct}{\mcitedefaultseppunct}\relax
\EndOfBibitem
\bibitem[Formanek and Moreno(2019)Formanek, and Moreno]{Formanek2019}
Formanek,~M.; Moreno,~A.~J. Single-Chain Nanoparticles under Homogeneous Shear
  Flow. \emph{Macromolecules} \textbf{2019}, \emph{52}, 1821--1831\relax
\mciteBstWouldAddEndPuncttrue
\mciteSetBstMidEndSepPunct{\mcitedefaultmidpunct}
{\mcitedefaultendpunct}{\mcitedefaultseppunct}\relax
\EndOfBibitem
\bibitem[Aust \latin{et~al.}(1999)Aust, Kr\"{o}ger, and Hess]{Aust1999}
Aust,~C.; Kr\"{o}ger,~M.; Hess,~S. Structure and Dynamics of Dilute Polymer
  Solutions under Shear Flow via Nonequilibrium Molecular Dynamics.
  \emph{Macromolecules} \textbf{1999}, \emph{32}, 5660--5672\relax
\mciteBstWouldAddEndPuncttrue
\mciteSetBstMidEndSepPunct{\mcitedefaultmidpunct}
{\mcitedefaultendpunct}{\mcitedefaultseppunct}\relax
\EndOfBibitem
\bibitem[Schroeder \latin{et~al.}(2005)Schroeder, Teixeira, Shaqfeh, and
  Chu]{Schroeder2005}
Schroeder,~C.~M.; Teixeira,~R.~E.; Shaqfeh,~E. S.~G.; Chu,~S. Dynamics of DNA
  in the Flow-Gradient Plane of Steady Shear Flow: Observations and
  Simulations. \emph{Macromolecules} \textbf{2005}, \emph{38}, 1967--1978\relax
\mciteBstWouldAddEndPuncttrue
\mciteSetBstMidEndSepPunct{\mcitedefaultmidpunct}
{\mcitedefaultendpunct}{\mcitedefaultseppunct}\relax
\EndOfBibitem
\bibitem[Ripoll \latin{et~al.}(2006)Ripoll, Winkler, and Gompper]{Ripoll2006}
Ripoll,~M.; Winkler,~R.~G.; Gompper,~G. Star Polymers in Shear Flow.
  \emph{Phys. Rev. Lett.} \textbf{2006}, \emph{96}, 188302\relax
\mciteBstWouldAddEndPuncttrue
\mciteSetBstMidEndSepPunct{\mcitedefaultmidpunct}
{\mcitedefaultendpunct}{\mcitedefaultseppunct}\relax
\EndOfBibitem
\bibitem[Nikoubashman and Likos(2010)Nikoubashman, and Likos]{Nikoubashman2010}
Nikoubashman,~A.; Likos,~C.~N. Branched Polymers under Shear.
  \emph{Macromolecules} \textbf{2010}, \emph{43}, 1610--1620\relax
\mciteBstWouldAddEndPuncttrue
\mciteSetBstMidEndSepPunct{\mcitedefaultmidpunct}
{\mcitedefaultendpunct}{\mcitedefaultseppunct}\relax
\EndOfBibitem
\bibitem[Chen \latin{et~al.}(2013)Chen, Chen, Liu, Xu, and An]{Chen2013}
Chen,~W.; Chen,~J.; Liu,~L.; Xu,~X.; An,~L. Effects of Chain Stiffness on
  Conformational and Dynamical Properties of Individual Ring Polymers in Shear
  Flow. \emph{Macromolecules} \textbf{2013}, \emph{46}, 7542--7549\relax
\mciteBstWouldAddEndPuncttrue
\mciteSetBstMidEndSepPunct{\mcitedefaultmidpunct}
{\mcitedefaultendpunct}{\mcitedefaultseppunct}\relax
\EndOfBibitem
\bibitem[Chen \latin{et~al.}(2013)Chen, Chen, and An]{Chen2013SM}
Chen,~W.; Chen,~J.; An,~L. Tumbling and tank-treading dynamics of individual
  ring polymers in shear flow. \emph{Soft Matter} \textbf{2013}, \emph{9},
  4312--4318\relax
\mciteBstWouldAddEndPuncttrue
\mciteSetBstMidEndSepPunct{\mcitedefaultmidpunct}
{\mcitedefaultendpunct}{\mcitedefaultseppunct}\relax
\EndOfBibitem
\bibitem[Chen \latin{et~al.}(2015)Chen, Li, Zhao, Liu, Chen, and An]{Chen2015}
Chen,~W.; Li,~Y.; Zhao,~H.; Liu,~L.; Chen,~J.; An,~L. Conformations and
  dynamics of single flexible ring polymers in simple shear flow.
  \emph{Polymer} \textbf{2015}, \emph{64}, 93 -- 99\relax
\mciteBstWouldAddEndPuncttrue
\mciteSetBstMidEndSepPunct{\mcitedefaultmidpunct}
{\mcitedefaultendpunct}{\mcitedefaultseppunct}\relax
\EndOfBibitem
\bibitem[Chen \latin{et~al.}(2017)Chen, Zhang, Liu, Chen, Li, and An]{Chen2017}
Chen,~W.; Zhang,~K.; Liu,~L.; Chen,~J.; Li,~Y.; An,~L. Conformation and
  Dynamics of Individual Star in Shear Flow and Comparison with Linear and Ring
  Polymers. \emph{Macromolecules} \textbf{2017}, \emph{50}, 1236--1244\relax
\mciteBstWouldAddEndPuncttrue
\mciteSetBstMidEndSepPunct{\mcitedefaultmidpunct}
{\mcitedefaultendpunct}{\mcitedefaultseppunct}\relax
\EndOfBibitem
\bibitem[Liebetreu \latin{et~al.}(2018)Liebetreu, Ripoll, and
  Likos]{Liebetreu2018}
Liebetreu,~M.; Ripoll,~M.; Likos,~C.~N. Trefoil Knot Hydrodynamic
  Delocalization on Sheared Ring Polymers. \emph{ACS Macro Lett.}
  \textbf{2018}, \emph{7}, 447--452\relax
\mciteBstWouldAddEndPuncttrue
\mciteSetBstMidEndSepPunct{\mcitedefaultmidpunct}
{\mcitedefaultendpunct}{\mcitedefaultseppunct}\relax
\EndOfBibitem
\bibitem[Jaramillo-Cano \latin{et~al.}(2018)Jaramillo-Cano, Formanek, Likos,
  and Camargo]{Jaramillo-Cano2018}
Jaramillo-Cano,~D.; Formanek,~M.; Likos,~C.~N.; Camargo,~M. Star
  Block-Copolymers in Shear Flow. \emph{J. Phys. Chem. B} \textbf{2018},
  \emph{122}, 4149--4158\relax
\mciteBstWouldAddEndPuncttrue
\mciteSetBstMidEndSepPunct{\mcitedefaultmidpunct}
{\mcitedefaultendpunct}{\mcitedefaultseppunct}\relax
\EndOfBibitem
\bibitem[Hur \latin{et~al.}(2001)Hur, Shaqfeh, Babcock, Smith, and
  Chu]{Hur2001}
Hur,~J.~S.; Shaqfeh,~E. S.~G.; Babcock,~H.~P.; Smith,~D.~E.; Chu,~S. Dynamics
  of dilute and semidilute DNA solutions in the start-up of shear flow.
  \emph{J. Rheol.} \textbf{2001}, \emph{45}, 421--450\relax
\mciteBstWouldAddEndPuncttrue
\mciteSetBstMidEndSepPunct{\mcitedefaultmidpunct}
{\mcitedefaultendpunct}{\mcitedefaultseppunct}\relax
\EndOfBibitem
\bibitem[Huang \latin{et~al.}(2010)Huang, Winkler, Sutmann, and
  Gompper]{Huang2010mac}
Huang,~C.-C.; Winkler,~R.~G.; Sutmann,~G.; Gompper,~G. Semidilute Polymer
  Solutions at Equilibrium and under Shear Flow. \emph{Macromolecules}
  \textbf{2010}, \emph{43}, 10107--10116\relax
\mciteBstWouldAddEndPuncttrue
\mciteSetBstMidEndSepPunct{\mcitedefaultmidpunct}
{\mcitedefaultendpunct}{\mcitedefaultseppunct}\relax
\EndOfBibitem
\bibitem[Huang \latin{et~al.}(2011)Huang, Sutmann, Gompper, and
  Winkler]{Huang2011}
Huang,~C.-C.; Sutmann,~G.; Gompper,~G.; Winkler,~R.~G. Tumbling of polymers in
  semidilute solution under shear flow. \emph{EPL} \textbf{2011}, \emph{93},
  54004\relax
\mciteBstWouldAddEndPuncttrue
\mciteSetBstMidEndSepPunct{\mcitedefaultmidpunct}
{\mcitedefaultendpunct}{\mcitedefaultseppunct}\relax
\EndOfBibitem
\bibitem[Huang \latin{et~al.}(2012)Huang, Gompper, and Winkler]{Huang2012jpcm}
Huang,~C.-C.; Gompper,~G.; Winkler,~R.~G. Non-equilibrium relaxation and
  tumbling times of polymers in semidilute solution. \emph{J. Phys.: Condens.
  Matter} \textbf{2012}, \emph{24}, 284131\relax
\mciteBstWouldAddEndPuncttrue
\mciteSetBstMidEndSepPunct{\mcitedefaultmidpunct}
{\mcitedefaultendpunct}{\mcitedefaultseppunct}\relax
\EndOfBibitem
\bibitem[Fedosov \latin{et~al.}(2012)Fedosov, Singh, Chatterji, Winkler, and
  Gompper]{Fedosov2012}
Fedosov,~D.~A.; Singh,~S.~P.; Chatterji,~A.; Winkler,~R.~G.; Gompper,~G.
  Semidilute solutions of ultra-soft colloids under shear flow. \emph{Soft
  Matter} \textbf{2012}, \emph{8}, 4109--4120\relax
\mciteBstWouldAddEndPuncttrue
\mciteSetBstMidEndSepPunct{\mcitedefaultmidpunct}
{\mcitedefaultendpunct}{\mcitedefaultseppunct}\relax
\EndOfBibitem
\bibitem[Singh \latin{et~al.}(2013)Singh, Chatterji, Gompper, and
  Winkler]{Singh2013}
Singh,~S.~P.; Chatterji,~A.; Gompper,~G.; Winkler,~R.~G. Dynamical and
  Rheological Properties of Ultrasoft Colloids under Shear Flow.
  \emph{Macromolecules} \textbf{2013}, \emph{46}, 8026--8036\relax
\mciteBstWouldAddEndPuncttrue
\mciteSetBstMidEndSepPunct{\mcitedefaultmidpunct}
{\mcitedefaultendpunct}{\mcitedefaultseppunct}\relax
\EndOfBibitem
\bibitem[Kremer and Grest(1990)Kremer, and Grest]{Kremer1990}
Kremer,~K.; Grest,~G.~S. Dynamics of entangled linear polymer melts: A
  molecular-dynamics simulation. \emph{{J}. {C}hem. {P}hys.} \textbf{1990},
  \emph{92}, 5057--5086\relax
\mciteBstWouldAddEndPuncttrue
\mciteSetBstMidEndSepPunct{\mcitedefaultmidpunct}
{\mcitedefaultendpunct}{\mcitedefaultseppunct}\relax
\EndOfBibitem
\bibitem[Lees and Edwards(1972)Lees, and Edwards]{Lees1972}
Lees,~A.; Edwards,~S. The computer study of transport processes under extreme
  conditions. \emph{J. Phys. C} \textbf{1972}, \emph{5}, 1921\relax
\mciteBstWouldAddEndPuncttrue
\mciteSetBstMidEndSepPunct{\mcitedefaultmidpunct}
{\mcitedefaultendpunct}{\mcitedefaultseppunct}\relax
\EndOfBibitem
\bibitem[Malevanets and Kapral(1999)Malevanets, and Kapral]{Malevanets1999}
Malevanets,~A.; Kapral,~R. Mesoscopic model for solvent dynamics. \emph{{J}.
  {C}hem. {P}hys.} \textbf{1999}, \emph{110}, 8605--8613\relax
\mciteBstWouldAddEndPuncttrue
\mciteSetBstMidEndSepPunct{\mcitedefaultmidpunct}
{\mcitedefaultendpunct}{\mcitedefaultseppunct}\relax
\EndOfBibitem
\bibitem[Everaers \latin{et~al.}(2004)Everaers, Sukumaran, Grest, Svaneborg,
  Sivasubramanian, and Kremer]{Ever_science}
Everaers,~R.; Sukumaran,~S.~K.; Grest,~G.~S.; Svaneborg,~C.;
  Sivasubramanian,~A.; Kremer,~K. Rheology and Microscopic Topology of
  Entangled Polymeric Liquids. \emph{Science} \textbf{2004}, \emph{303},
  823--826\relax
\mciteBstWouldAddEndPuncttrue
\mciteSetBstMidEndSepPunct{\mcitedefaultmidpunct}
{\mcitedefaultendpunct}{\mcitedefaultseppunct}\relax
\EndOfBibitem
\bibitem[Hoy \latin{et~al.}(2009)Hoy, Foteinopoulou, and Kr\"oger]{Kroger_Ne}
Hoy,~R.~S.; Foteinopoulou,~K.; Kr\"oger,~M. Topological analysis of polymeric
  melts: Chain-length effects and fast-converging estimators for entanglement
  length. \emph{Phys. Rev. E} \textbf{2009}, \emph{80}, 031803\relax
\mciteBstWouldAddEndPuncttrue
\mciteSetBstMidEndSepPunct{\mcitedefaultmidpunct}
{\mcitedefaultendpunct}{\mcitedefaultseppunct}\relax
\EndOfBibitem
\bibitem[Arbe \latin{et~al.}(2019)Arbe, Rubio-Cervilla, Alegr\'{\i}a, Moreno,
  Pomposo, Robles-Hern\'{a}ndez, Malo~de Molina, Fouquet, Juranyi, and
  Colmenero]{Arbe2019meso}
Arbe,~A.; Rubio-Cervilla,~J.; Alegr\'{\i}a,~A.; Moreno,~A.~J.; Pomposo,~J.~A.;
  Robles-Hern\'{a}ndez,~B.; Malo~de Molina,~P.; Fouquet,~P.; Juranyi,~F.;
  Colmenero,~J. Mesoscale Dynamics in Melts of Single-Chain Polymeric
  Nanoparticles. \emph{Macromolecules} \textbf{2019}, \emph{52},
  6935--6942\relax
\mciteBstWouldAddEndPuncttrue
\mciteSetBstMidEndSepPunct{\mcitedefaultmidpunct}
{\mcitedefaultendpunct}{\mcitedefaultseppunct}\relax
\EndOfBibitem
\bibitem[Dalal \latin{et~al.}(2012)Dalal, Albaugh, Hoda, and Larson]{Dalal2012}
Dalal,~S.~I.; Albaugh,~A.; Hoda,~N.; Larson,~R.~G. Tumbling and Deformation of
  Isolated Polymer Chains in Shearing Flow. \emph{Macromolecules}
  \textbf{2012}, \emph{45}, 9493--9499\relax
\mciteBstWouldAddEndPuncttrue
\mciteSetBstMidEndSepPunct{\mcitedefaultmidpunct}
{\mcitedefaultendpunct}{\mcitedefaultseppunct}\relax
\EndOfBibitem
\bibitem[Lang \latin{et~al.}(2014)Lang, Obermayer, and Frey]{Lang2014}
Lang,~P.~S.; Obermayer,~B.; Frey,~E. Dynamics of a semiflexible polymer or
  polymer ring in shear flow. \emph{Phys. Rev. E} \textbf{2014}, \emph{89},
  022606\relax
\mciteBstWouldAddEndPuncttrue
\mciteSetBstMidEndSepPunct{\mcitedefaultmidpunct}
{\mcitedefaultendpunct}{\mcitedefaultseppunct}\relax
\EndOfBibitem
\bibitem[Bossart and Oettinger(1995)Bossart, and Oettinger]{Bossart1995}
Bossart,~J.; Oettinger,~H.~C. Orientation of Polymer Coils in Dilute Solutions
  Undergoing Shear Flow: Birefringence and Light Scattering.
  \emph{Macromolecules} \textbf{1995}, \emph{28}, 5852--5860\relax
\mciteBstWouldAddEndPuncttrue
\mciteSetBstMidEndSepPunct{\mcitedefaultmidpunct}
{\mcitedefaultendpunct}{\mcitedefaultseppunct}\relax
\EndOfBibitem
\bibitem[Bird \latin{et~al.}(1987)Bird, Curtiss, Armstrong, and
  Hassager]{Bird1987}
Bird,~R.; Curtiss,~C.; Armstrong,~R.; Hassager,~O. \emph{Dynamics of Polymer
  Liquids Vol. 2 Kinetic Theory}; Wiley, 1987\relax
\mciteBstWouldAddEndPuncttrue
\mciteSetBstMidEndSepPunct{\mcitedefaultmidpunct}
{\mcitedefaultendpunct}{\mcitedefaultseppunct}\relax
\EndOfBibitem
\bibitem[Bekard \latin{et~al.}(2011)Bekard, Asimakis, Bertolini, and
  Dunstan]{Bekard2011}
Bekard,~I.~B.; Asimakis,~P.; Bertolini,~J.; Dunstan,~D.~E. The effects of shear
  flow on protein structure and function. \emph{Biopolymers} \textbf{2011},
  \emph{95}, 733--745\relax
\mciteBstWouldAddEndPuncttrue
\mciteSetBstMidEndSepPunct{\mcitedefaultmidpunct}
{\mcitedefaultendpunct}{\mcitedefaultseppunct}\relax
\EndOfBibitem
\end{mcitethebibliography}

\begin{mcitethebibliography}{13}
\providecommand*\natexlab[1]{#1}
\providecommand*\mciteSetBstSublistMode[1]{}
\providecommand*\mciteSetBstMaxWidthForm[2]{}
\providecommand*\mciteBstWouldAddEndPuncttrue
  {\def\EndOfBibitem{\unskip.}}
\providecommand*\mciteBstWouldAddEndPunctfalse
  {\let\EndOfBibitem\relax}
\providecommand*\mciteSetBstMidEndSepPunct[3]{}
\providecommand*\mciteSetBstSublistLabelBeginEnd[3]{}
\providecommand*\EndOfBibitem{}
\mciteSetBstSublistMode{f}
\mciteSetBstMaxWidthForm{subitem}{(\alph{mcitesubitemcount})}
\mciteSetBstSublistLabelBeginEnd
  {\mcitemaxwidthsubitemform\space}
  {\relax}
  {\relax}

\bibitem[Kremer and Grest(1990)Kremer, and Grest]{Kremer1990bis}
Kremer,~K.; Grest,~G.~S. Dynamics of entangled linear polymer melts: A
  molecular-dynamics simulation. \emph{{J}. {C}hem. {P}hys.} \textbf{1990},
  \emph{92}, 5057--5086\relax
\mciteBstWouldAddEndPuncttrue
\mciteSetBstMidEndSepPunct{\mcitedefaultmidpunct}
{\mcitedefaultendpunct}{\mcitedefaultseppunct}\relax
\EndOfBibitem
\bibitem[Izaguirre \latin{et~al.}(2001)Izaguirre, Catarello, Wozniak, and
  Skeel]{Izaguirre2001}
Izaguirre,~J.~A.; Catarello,~D.~P.; Wozniak,~J.~M.; Skeel,~R.~D. Langevin
  stabilization of molecular dynamics. \emph{{J} {C}hem. {P}hys.}
  \textbf{2001}, \emph{114}, 2090--2098\relax
\mciteBstWouldAddEndPuncttrue
\mciteSetBstMidEndSepPunct{\mcitedefaultmidpunct}
{\mcitedefaultendpunct}{\mcitedefaultseppunct}\relax
\EndOfBibitem
\bibitem[Formanek and Moreno(2019)Formanek, and Moreno]{Formanek2019bis}
Formanek,~M.; Moreno,~A.~J. Single-Chain Nanoparticles under Homogeneous Shear
  Flow. \emph{Macromolecules} \textbf{2019}, \emph{52}, 1821--1831\relax
\mciteBstWouldAddEndPuncttrue
\mciteSetBstMidEndSepPunct{\mcitedefaultmidpunct}
{\mcitedefaultendpunct}{\mcitedefaultseppunct}\relax
\EndOfBibitem
\bibitem[Malevanets and Kapral(1999)Malevanets, and Kapral]{Malevanets1999bis}
Malevanets,~A.; Kapral,~R. Mesoscopic model for solvent dynamics. \emph{{J}.
  {C}hem. {P}hys.} \textbf{1999}, \emph{110}, 8605--8613\relax
\mciteBstWouldAddEndPuncttrue
\mciteSetBstMidEndSepPunct{\mcitedefaultmidpunct}
{\mcitedefaultendpunct}{\mcitedefaultseppunct}\relax
\EndOfBibitem
\bibitem[Malevanets and Kapral(2000)Malevanets, and Kapral]{Malevanets2000}
Malevanets,~A.; Kapral,~R. Solute molecular dynamics in a mesoscale solvent.
  \emph{J. Chem. Phys.} \textbf{2000}, \emph{112}, 7260--7269\relax
\mciteBstWouldAddEndPuncttrue
\mciteSetBstMidEndSepPunct{\mcitedefaultmidpunct}
{\mcitedefaultendpunct}{\mcitedefaultseppunct}\relax
\EndOfBibitem
\bibitem[Ihle and Kroll(2001)Ihle, and Kroll]{Ihle2001}
Ihle,~T.; Kroll,~D. Stochastic rotation dynamics: a Galilean-invariant
  mesoscopic model for fluid flow. \emph{Phys. Rev. E} \textbf{2001},
  \emph{63}, 020201\relax
\mciteBstWouldAddEndPuncttrue
\mciteSetBstMidEndSepPunct{\mcitedefaultmidpunct}
{\mcitedefaultendpunct}{\mcitedefaultseppunct}\relax
\EndOfBibitem
\bibitem[Ihle and Kroll(2003)Ihle, and Kroll]{Ihle2003}
Ihle,~T.; Kroll,~D.~M. Stochastic rotation dynamics. I. Formalism, Galilean
  invariance, and Green-Kubo relations. \emph{Phys. Rev. E} \textbf{2003},
  \emph{67}, 066705\relax
\mciteBstWouldAddEndPuncttrue
\mciteSetBstMidEndSepPunct{\mcitedefaultmidpunct}
{\mcitedefaultendpunct}{\mcitedefaultseppunct}\relax
\EndOfBibitem
\bibitem[Lees and Edwards(1972)Lees, and Edwards]{Lees1972bis}
Lees,~A.; Edwards,~S. The computer study of transport processes under extreme
  conditions. \emph{J. Phys. C} \textbf{1972}, \emph{5}, 1921\relax
\mciteBstWouldAddEndPuncttrue
\mciteSetBstMidEndSepPunct{\mcitedefaultmidpunct}
{\mcitedefaultendpunct}{\mcitedefaultseppunct}\relax
\EndOfBibitem
\bibitem[Mussawisade \latin{et~al.}(2005)Mussawisade, Ripoll, Winkler, and
  Gompper]{Mussawisade2005}
Mussawisade,~K.; Ripoll,~M.; Winkler,~R.~G.; Gompper,~G. Dynamics of polymers
  in a particle-based mesoscopic solvent. \emph{J. Chem. Phys.} \textbf{2005},
  \emph{123}\relax
\mciteBstWouldAddEndPuncttrue
\mciteSetBstMidEndSepPunct{\mcitedefaultmidpunct}
{\mcitedefaultendpunct}{\mcitedefaultseppunct}\relax
\EndOfBibitem
\bibitem[Frenkel and Smit(1996)Frenkel, and Smit]{Frenkel1996}
Frenkel,~D.; Smit,~B. \emph{Understanding molecular simulations: from
  algorithms to applications}; Academic Press, 1996\relax
\mciteBstWouldAddEndPuncttrue
\mciteSetBstMidEndSepPunct{\mcitedefaultmidpunct}
{\mcitedefaultendpunct}{\mcitedefaultseppunct}\relax
\EndOfBibitem
\bibitem[Singh \latin{et~al.}(2014)Singh, Huang, Westphal, Gompper, and
  Winkler]{Singh2014}
Singh,~S.~P.; Huang,~C.-C.; Westphal,~E.; Gompper,~G.; Winkler,~R.~G.
  Hydrodynamic correlations and diffusion coefficient of star polymers in
  solution. \emph{J. Chem. Phys.} \textbf{2014}, \emph{141}, 084901\relax
\mciteBstWouldAddEndPuncttrue
\mciteSetBstMidEndSepPunct{\mcitedefaultmidpunct}
{\mcitedefaultendpunct}{\mcitedefaultseppunct}\relax
\EndOfBibitem
\bibitem[Huang \latin{et~al.}(2010)Huang, Winkler, Sutmann, and
  Gompper]{Huang2010macbis}
Huang,~C.-C.; Winkler,~R.~G.; Sutmann,~G.; Gompper,~G. Semidilute Polymer
  Solutions at Equilibrium and under Shear Flow. \emph{Macromolecules}
  \textbf{2010}, \emph{43}, 10107--10116\relax
\mciteBstWouldAddEndPuncttrue
\mciteSetBstMidEndSepPunct{\mcitedefaultmidpunct}
{\mcitedefaultendpunct}{\mcitedefaultseppunct}\relax
\EndOfBibitem
\end{mcitethebibliography}
\end{document}